\documentclass{article}

\usepackage[eandd,final]{neurips_2026}

\usepackage[utf8]{inputenc} % allow utf-8 input
\usepackage[T1]{fontenc}    % use 8-bit T1 fonts
\usepackage{hyperref}       % hyperlinks
\usepackage{url}            % simple URL typesetting
\usepackage{booktabs}       % professional-quality tables
\usepackage{amsfonts}       % blackboard math symbols
\usepackage{nicefrac}       % compact symbols for 1/2, etc.
\usepackage{microtype}      % microtypography
\usepackage{xcolor}         % colors
\usepackage{dirtree}
\usepackage{caption}
\usepackage{wrapfig}
\usepackage{tabularx}
\usepackage{booktabs}
\usepackage{natbib}
\usepackage{graphicx}
\usepackage{multirow}
\usepackage{amsmath} 
\usepackage{array}
\usepackage{float}

\title{L-FAME: Longitudinal Focused Attention Meditation EEG Dataset and Benchmark}

\author{%
  Angqi Li \\
  Department of CMSE \\
  Michigan State University \\
  East Lansing, MI 48824 \\
  \texttt{liangqi1@msu.edu} \\
  \And
  Ab Basit Rafi Syed \\
  Department of CMSE \\
  Michigan State University \\
  East Lansing, MI 48824 \\
  \texttt{syedab@msu.edu} \\
  \And
  Hamzeh Alzweri \\
   Department of CSE \\
  Michigan State University \\
  East Lansing, MI 48824 \\
  \texttt{alzwerih@msu.edu} \\
  \AND
  Taosheng Liu \\
  Department of Psychology \\
  Michigan State University \\
  East Lansing, MI 48824 \\
  \texttt{tsliu@msu.edu} \\
  \And
  Barry H.~Cohen \\
  Department of Applied Psychology \\
  New York University \\
  New York, NY 10003 \\
  \texttt{barry.cohen@nyu.edu} \\
  \And
  Saiprasad Ravishankar \\
  Department of CMSE \\
  Department of BME \\
  Michigan State University \\
  East Lansing, MI 48824 \\
  \texttt{ravisha3@msu.edu} \\
}

\begin{document}

\maketitle

\begin{abstract}
We introduce a novel Longitudinal Focused Attention Meditation Electroencephalography (L-FAME) dataset and an accompanying benchmark, designed to foster research into the %distinct 
neural effects of various meditation practices and the evolution of these effects over a six-week training period. The dataset includes baseline EEG recordings from 74 healthy college participants, of whom 44 completed the six-week intervention and returned for post-intervention recordings, along with psychological assessments. Participants were randomly assigned to one of three distinct meditation groups: two mantra-based techniques (SA-TA-NA-MA and Hare Krishna) and one Breath Focus practice. Leveraging this unique longitudinal and comparative dataset, we propose a benchmark suite comprising three distinct classification tasks: (1) cognitive state decoding to distinguish between resting and meditation states, (2) fine-grained classification of the specific meditation techniques, and (3) cross-session adaptation to evaluate model generalization across the longitudinal time gap. We provide comprehensive baseline results for these tasks utilizing a range of classical machine learning algorithms and deep learning architectures. The complete dataset, preprocessing pipelines, and benchmark evaluation code are being publicly released, offering a valuable resource and a standardized framework for the development and comparison of new analytical methods in computational meditation research and EEG-based machine learning. \url{https://huggingface.co/datasets/L-FAME-Dataset-Benchmark/L-FAME}
\end{abstract}

\section{Introduction}

Brain-computer interfaces (BCIs) and neural decoding algorithms have received substantial attention as they transition into practical applications. Electroencephalography (EEG) remains the preferred modality for these tasks due to its non-invasive nature, high temporal resolution, cost-effectiveness, and portability~\cite{chaddad2023electroencephalography}. These strengths, coupled with advancements in machine learning and deep learning architectures tailored for neural signals, have enabled breakthroughs in motor imagery and stimulus-evoked decoding~\cite{bel2022scoping,logie2003brain}. Beyond these traditional tasks, EEG is increasingly utilized to decode internally generated or spontaneous continuous cognitive states, such as sustained attention, mental fatigue, and mind-wandering. However, unlike stimulus-evoked paradigms, these spontaneous states lack explicit external triggers, which makes precise objective labeling exceptionally challenging~\cite{fan2025deep,liu2017automated,tang2023mind}. Consequently, critical deficiencies remain in the definition of tasks and the standardization of protocols. 

Focused attention meditation (FAM) serves as both a continuous cognitive task and a clinical therapeutic tool for mental health. 
Research indicates that the neurobehavioral signatures of FAM are distinct from those of mind-wandering~\cite{rodriguez2024assessing,zhang2021longitudinal,li2026not}. This distinction positions FAM as an ideal experimental proxy for evaluating whether deep learning models can capture subtle, non-stationary changes in neural topology. Furthermore, algorithmic developments in this area can support clinical meditation training and the quality control of interventions~\cite{burgess2021brief,khoury2015mindfulness}.

% \textcolor{blue}{Sai: add some references here for clinical use of meditation.}

The recent integration of deep learning with EEG analysis provides promising methods for objectively quantifying meditative states and facilitating personalized mental health monitoring~\cite{lyu2026deep,liu2025leveraging}. However, the practical deployment of these models requires robust generalization over time. Most existing datasets and studies on the decoding of meditation remain cross-sectional~\cite{brandmeyer2019neuroscience,novelli2025psiconnect}. While comparisons between meditators and non-meditators yield valuable insights, they cannot account for the dynamic changes within individuals during longitudinal training over weeks or months. Models trained on single-session data may achieve high accuracy in intra-subject evaluations, but they often experience catastrophic performance degradation when processing data from the same participant weeks later~\cite{tang2015neuroscience,jayaram2018moabb,lotte2007review}. This degradation is primarily due to non-stationary temporal shifts. Furthermore, existing longitudinal datasets are often too small to support robust cross-subject generalization~\cite{shang2023eeg}. 

To address the gaps in temporal and cross-subject distribution shifts, we introduce the Longitudinal Focused Attention Meditation EEG (L-FAME) dataset. This dataset comprises high-resolution 64-channel EEG recordings from an initial cohort of 74 participants. It includes a subset of 44 individuals who completed a six-week pre- and post-intervention protocol. Rather than focusing on a single tradition, L-FAME systematically compares three distinct FAM paradigms, namely Breath Focus (BF), Hare Krishna (HK), and SA-TA-NA-MA (SA)~\cite{lutz2008attention}. 
We establish a benchmark suite that ranges from state classification to fine-grained technique classification and transfer learning. These tasks specifically assess cross-session generalization and cross-subject domain adaptation. They provide a standardized framework to evaluate how deep learning models manage non-stationary neural drifts over time.

The contributions of this work are summarized as follows:

\begin{itemize}

   \item \textbf{A High-Quality Longitudinal EEG Dataset:} A well-documented, preprocessed, and publicly accessible 64-channel EEG dataset spanning three FAM practices. The dataset includes 74 participants at baseline, of whom 44 completed the six-week intervention and both pre- and post-intervention EEG recordings.
    
    \item \textbf{Novel Benchmark Suite:} We present a benchmark machine learning and deep learning (ML \& DL) framework consisting of three evaluation tasks applied to a subset of our longitudinal meditation EEG dataset.

    \item \textbf{Psychometric Behavioral Labels:} We also provide psychological assessments collected at two time points to establish a behavioral ground truth for changes related to meditation.
    \item \textbf{Bridging Disciplines:} Beyond developing ML/DL for neural decoding, this dataset serves as a vital resource for the neuroscience community to investigate the neurophysiological mechanisms underlying meditation.

\end{itemize}

\vspace{0.2in}
\section{Related Work}

Recent advances in EEG decoding have expanded beyond stimulus-locked tasks~\cite{nicolas2012brain,jayaram2018moabb} to continuous cognitive states, such as mental fatigue, affective states, and sustained attention~\cite{koelstra2011deap,zheng2015investigating}. Evaluating these continuous states is crucial for the real-world deployment of brain-computer interfaces (BCIs), which require machine learning models to capture spontaneous, non-stationary neural dynamics over extended periods without explicit external triggers~\cite{huang2023discrepancy,krusienski2011critical}. Within this broader landscape, decoding meditative states, specifically differentiating FAM from spontaneous mind-wandering (MW), represents a unique and challenging frontier~\cite{shang2023eeg}. Unlike passive emotional or sensory responses, meditation involves active internal self-regulation. Consequently, this process serves as an ideal paradigm for testing the ability of a model to track subtle, purely internally driven temporal shifts. Although recent studies have begun to explore deep learning for meditation decoding, progress remains heavily constrained by dataset limitations.
%\textcolor{blue}{Sai: Make sure no citations are appearing as question marks. I so wish to see natural human sentences. Every paper seems like GPT based now. Need to include Prof. GPT as author also :)}

\paragraph{Related Datasets.}
Several publicly available EEG datasets incorporate meditation or related cognitive tasks. However, compared to L-FAME, they are often acquired using cross-sectional designs with smaller cohorts~\cite{brandmeyer2018reduced} or focus exclusively on specific sub-populations, like highly experienced monks~\cite{Wongupparaj2024EEGab}, which limits the ability to evaluate training-induced temporal dynamics in the general population. A prominent example of a large-scale multimodal longitudinal dataset is PsiConnect, including EEG and fMRI data from a significant sample of participants, with a subset of 30 individuals completing an 8-week mindfulness program~\cite{novelli2025psiconnect}. However, it primarily investigates the interaction between psilocybin and meditation, introducing complex pharmacological variables that may fall outside the scope of pure cognitive state decoding. While some recent pure longitudinal tracking efforts are severely bottlenecked by small sample sizes and single-technique restrictions~\cite{shang2023eeg}, L-FAME provides a uniquely rigorous resource, by offering high-resolution recordings, robust longitudinal cohorts, and a systematic comparison of multiple FAM paradigms. Table~\ref{tab:meditation datasets} presents a comparative summary of L-FAME and other public EEG datasets in this domain.% \textcolor{blue}{Sai: incomplete sentence}

\renewcommand{\arraystretch}{1.25}
\begin{table}[h]
\centering
\caption{\small Comparison of the L-FAME dataset with existing publicly available EEG datasets for meditation.}
\label{tab:meditation datasets}
\begin{tabular}{@{} l p{3.7cm} c c p{3.2cm} @{}}
\toprule
Dataset & Meditation Type(s) & N & Age & Design \\
\midrule
Delorme \& Brandmeyer~\cite{brandmeyer2018reduced} & Focused Attention & 24 & 31--78 & Cross sectional \\
Wongupparaj et al.~\cite{Wongupparaj2024EEGab} & Mindfulness & 60 & 19--70 & Cross sectional \\
PsiConnect~\cite{novelli2025psiconnect}& Mindfulness & 62$^*$ & 18--55 & 8-week intervention \\
Shang et al.~\cite{shang2023eeg} & Mindfulness-Based Stress Reduction (MBSR) & 11 & 25-45 & 6-week intervention \\
\textbf{L-FAME (Ours)} & \textbf{Focused Attention$^\dagger$} & \textbf{74} & \textbf{18--32} & \textbf{6-week intervention} \\
\bottomrule
\multicolumn{5}{l}{\scriptsize $^*$ Only a subset of 30 participants underwent the meditation training.  $^\dagger$ Three different FAM techniques.}
\end{tabular}
\end{table}

\vspace{0.2in}
\section{L-FAME Dataset}

\subsection{Data Acquisition}

\paragraph{Participants.} A total of 74 healthy college students (mean age: $22$ years; 46 females, 28 males) were recruited from Michigan State University to participate in the study. The cohort consisted primarily of right-handed individuals (68 right-handed, 5 left-handed, 1 missing). Following recruitment, participants were randomly assigned to one of three meditation training groups: Breath Focus (BF), Hare Krishna (HK), or SA-TA-NA-MA (SA) . Collected demographic data included age, sex, and handedness. A comprehensive summary of the demographics of the participants is provided in Table~\ref{tab:participant-demographics} and the detailed participant demographics can be found in Appendix~\ref{Detailed_Participant}. Detailed criteria for screening are in Appendix~\ref{Criteria} and descriptions of the specific meditation techniques appear in Appendix~\ref{techniques}.

\begin{table}[h]
  \centering
  \caption{Participant demographics by meditation paradigm ($N=74$). All paradigms follow the Focused Attention (FA) framework with distinct cognitive focus objects.}
  \label{tab:participant-demographics}
  \begin{tabular}{@{} l l l c c c @{}}
    \toprule
    Category & Paradigm & Focus Object & N & Female/Male & Age (mean ± SD) \\
    \midrule
    \multirow{3}{*}{\shortstack[l]{Focused \\ Attention \\Meditation \\ (FAM)}} 
      & Breath Focus (BF) & Respiration & 16 & 11 / 5 & 22.2 ± 3.9 \\
      & Hare Krishna (HK) & Long mantra & 31 & 18 / 13 & 22.2 ± 4.2 \\
      & SA-TA-NA-MA (SA)  & Simpler mantra & 27 & 17 / 10 & 21.7 ± 2.7 \\
    \midrule
    \textbf{Total} & & & \textbf{74} & \textbf{46 / 28} & \textbf{22.0 ± 3.6} \\
    \bottomrule
  \end{tabular}
\end{table}

\paragraph{Interventions.} Participants were trained in specific techniques categorized under FAM. These variants included focusing on sensations created by breathing, referred to here as Breath Focus, and Japa (mantra-based meditation), which utilized the mantras SA-TA-NA-MA and Hare Krishna.

% \textcolor{blue}{Sai: Didn't we group Japa as FAM. Better to make above paragraph precise (addressed)}

\paragraph{EEG Recording setup.}EEG data were recorded in a controlled environment using a \textit{mBrainTrain Smarting Pro X} amplifier (mBrainTrain, Belgrade, Serbia) and a 64-channel cap manufactured by \textit{EASYCAP} (EASYCAP, Hersching, Germany). The cap featured Ag/AgCl electrodes arranged according to the international 10-10 system layout. To ensure high signal fidelity, the FCz electrode served as the reference, while FPz was utilized as the ground. High-chloride abrasive electrolyte gel (abralyt HiCl gel) was applied to achieve stable conductivity, with electrode impedances strictly maintained below 20 $k\Omega$ throughout the session.

\subsection{Experimental Paradigms}

\begin{figure}
    \centering
    \includegraphics[width=\linewidth]{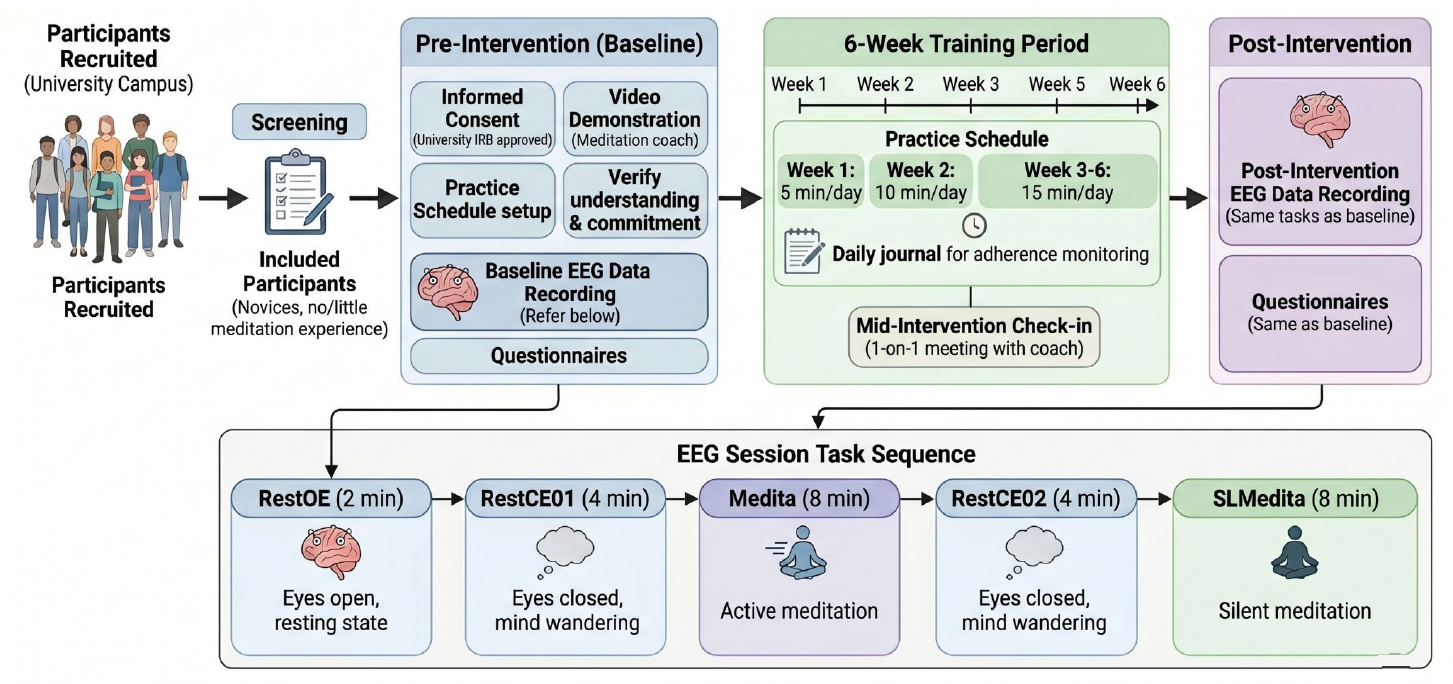}
    \caption{\small Overview of Longitudinal Meditation Study Design and EEG Task Sequence}
    \label{fig:study design}
\end{figure}

In this section, we provide an overview of the experimental paradigms used in our study, which consists of a structured longitudinal design and a standardized EEG session protocol. Participants were screened, as detailed in Appendix~\ref{Criteria}, and were guided through a consistent daily practice schedule, detailed in Appendix~\ref{techniques}. EEG sessions conducted before and after the 6-week intervention included a sequence of resting and meditation tasks, allowing for evaluation of both state- and trait-level neural changes associated with different meditation practices. The full design is shown in Figure~\ref{fig:study design}.

\subsubsection{Longitudinal Design}

% \begin{wrapfigure}{r}{0.4\textwidth}
%   \centering
%   \includegraphics[width=0.4\textwidth]{figures/Longitudinal Design.png}  
%   \caption{Flow chart of longitudinal design}
%   \label{fig:wrapped}
% \end{wrapfigure}

% \paragraph{Screening.} Potential adult novice participants were recruited primarily from Michigan State University (MSU) campus and underwent an initial screening process by answering a questionnaire (see Appendix~\ref{Criteria}) This ensured participants had no or little prior meditation experience.

\paragraph{Screening.} Potential adult novice participants were recruited primarily from Michigan State University (MSU) campus and underwent an initial screening process based on specific inclusion and exclusion criteria (see Section~\ref{Criteria}). This ensured participants had little to no prior meditation experience.

\paragraph{Pre-Intervention (Baseline).} Following successful screening and before the start of the 6-week training period, participants attended a baseline session. Upon arrival, they first read and signed the MSU Institutional Review Board (IRB) approved consent form. Next, participants were shown a pre-recorded video by the study's meditation coach demonstrating the meditation practice corresponding to their assigned group. Researchers then verified that the participants understood the practice and could commit to the daily schedule. Finally, the participants completed all baseline data collection, including an EEG recording session detailed in Section~\ref{eegtasks} and questionnaires (see Appendix~\ref{questionnaires}). After all tests are done, participants were instructed to practice daily following a gradually increasing schedule (see Appendix~\ref{techniques}).

\paragraph{Mid-Intervention (Check-in).} After three weeks of practice, participants scheduled a one-on-one meeting with the meditation coach to discuss their progress and address any questions.

\paragraph{Post-Intervention.} Within one week following the completion of the 6-week training period, participants returned for a post-intervention session. During this session, they performed the same tasks as in the baseline, including the EEG data recording detailed in Section~\ref{eegtasks}.

\subsubsection{EEG Session Tasks}
\label{eegtasks}

During both the baseline and post-intervention EEG sessions, participants completed a standardized sequence of five tasks. The protocol comprised a two-minute resting state with eyes open (restOE), followed by a four-minute resting state with eyes closed (restCE01) during which participants were instructed to allow their mind to  wander. Subsequently, participants engaged in an eight-minute active meditation (Medita). Because meditative practices frequently utilize overt physical or vocal engagement to facilitate initial focus, this active phase was incorporated despite the introduction of artifacts. Following a second four-minute resting state (restCE02) utilizing identical instructions, the session concluded with an eight-minute silent meditation (slMedita). During this final phase, participants executed the assigned techniques entirely internally without movement. This silent execution demands greater cognitive focus while providing the reducing movement- and vocalization-related artifacts neural data required for the primary analyses. Instructions were provided orally by the research assistant; comprehensive details regarding the specific task execution and the synchronization of event markers are available in Appendix~\ref{EEG session detail}.

\subsection{Data Structure and Format}
\label{data_struct}

The dataset is organized and formatted in accordance with the Brain Imaging Data Structure (BIDS) standard, specifically the BIDS-EEG extension~\cite{gorgolewski2016brain}. Adherence to BIDS ensures standardization, facilitates data sharing and reuse, and promotes interoperability with a diverse range of analysis software, e.g., MNE-Python, EEGLAB, and FieldTrip. Following the BIDS guidelines, each data file is accompanied by a corresponding metadata file in JSON format.

\begin{table}[ht]
    \centering
    \small
    \caption{Overview of dataset structure and processing derivatives.}
    \begin{tabular}{lllll}
        \toprule
        Data type & Folder & Participant ID & Session & Modality \\
        \midrule
        Defaced raw data & /L-FAME & \begin{tabular}[c]{@{}l@{}}/sub-xx \end{tabular} & \begin{tabular}[c]{@{}l@{}}/ses-premedita\\ /ses-posmedita\end{tabular} & \begin{tabular}[c]{@{}l@{}}BrainVision \\(.eeg, .vhdr, .vmrk) \end{tabular} \\
        \midrule
        Training data & ./derivatives/ml\_preproc\_data & \begin{tabular}[c]{@{}l@{}}/sub-xxx \end{tabular} & N/A  & \begin{tabular}[c]{@{}l@{}}Tensors (.npy) \end{tabular} \\
        \midrule
        Cleaned EEG & ./derivatives/eeglab\_preproc & \begin{tabular}[c]{@{}l@{}}/sub-xxx \end{tabular} & \begin{tabular}[c]{@{}l@{}}/ses-premedita\\ /ses-posmedita\end{tabular} & \begin{tabular}[c]{@{}l@{}} EEGlab (.set) \end{tabular} \\
        \bottomrule
    \end{tabular}
\end{table}

The de-identified raw EEG data are provided in BrainVision format, with all supplementary information, such as electrode locations and event markers, complying with the BIDS guidelines. Furthermore, the continuous data are pre-segmented into five tasks and saved in the EEGLAB format. This structure allows researchers to apply any preferred preprocessing methods for subsequent analysis. In addition, a cleaned EEG dataset is provided, which underwent the preprocessing pipeline described below for immediate further analysis. Detailed preprocessing steps 
%for following paragraphs 
are included in Appendix~\ref{data_preprocess}.

\paragraph{Quantitative Quality Assessment.} We assessed raw-recording quality separately from benchmark-input quality. Archived quality check (QC) results for 118 raw recording sessions showed that a median of 98.4\% of channels met the QC criteria before interpolation; three sessions (2.54\%) had fewer than 90\% of channels meeting these criteria. The median four-neighbour channel correlation was 0.884. We additionally assessed 472 minimally preprocessed task segments, each corresponding to one participant's recording of one task within a session. Within each task segment, we checked each channel in non-overlapping 4-s windows for threshold-exceeding amplitude or low variability and calculated the proportion of checks that were flagged. The median proportion across task segments was 0.56\%, indicating that these threshold violations were uncommon in a typical task recording, although some recordings had substantially higher flagged proportions. Together, these results support generally high recording quality under the assessed QC criteria. QC criteria, metric definitions, and exceptions are detailed in Appendix~\ref{app:eeg-qc}.

\paragraph{Cleaned EEG.} Meditation recordings were partitioned into five segments defined in Section~\ref{eegtasks}. To ensure signal quality for machine learning, we implemented a preprocessing pipeline focusing on automated artifact suppression and signal retention. This included zero-phase high-pass filtering \citep{WidmannSchrogerMaess2015}, Zapline-plus line-noise removal \citep{klug2022zaplineplus}, and Artifact Subspace Reconstruction (ASR) \citep{Mullen2012}. Neural activity was isolated via Independent Component Analysis (ICA) with ICLabel classification \citep{PionTonachiniKreutzDelgadoMakeig2019}. We provide both pre-ICA and IC-removed versions (extensions -preica and -icrm) to support diverse research requirements.

% \textcolor{blue}{(addressed) Sai: Training Data paragraph was confusing to me. Rewrite to explain clearly. }

\paragraph{Training Data.} This subset establishes benchmark tasks' training and testing data for ML/DL following established preprocessing protocols \cite{delorme2023eeg, kessler2025eeg}. Processing involved a 0.5 Hz FIR high-pass filter to attenuate low-frequency drifts and automated bad-channel identification via the eeglab algorithm (`clean\_raw'). Spherical spline interpolation was applied to ensure uniform spatial dimensions across all recordings.

\vspace{0.2in}

\section{Benchmark}

We define three core benchmark tasks (Task 1, Task 2, and Task 3) to evaluate model capacity and generalization on the L-FAME dataset. Continuous EEG recordings are segmented into four-second samples using sliding windows. We use a $50\%$ overlap for cross-subject evaluations and $75\%$ for intra-subject evaluations to augment data samples and ensure model convergence.Three progressive evaluation strategies measure generalization. Intra-subject evaluation measures individual-level stability via a block-wise interleaved split to mitigate temporal concept drift. Inter-subject evaluation employs a 5-fold cross-validation strictly partitioned at the subject level. In each fold, models are trained on data from $80\%$ of the subjects and evaluated on the remaining $20\%$ of completely unseen subjects, preventing data leakage and rigorously assessing population-level generalization. Leave-one-subject-out (LOSO) cross-validation provides a strict test for subject-independent generalization.This framework isolates performance aspects: intra-subject evaluation establishes a theoretical upper bound, inter-subject evaluation offers a baseline for model fine-tuning, and LOSO simulates clinical scenarios with unseen subjects to expose inter-individual variability. A dynamic weighted random sampler ensures class balance; detailed technical specifications for the block-wise interleaved split, sampling, and benchmark tasks are provided in Appendices~\ref{training_detail} and \ref{benchmark_detail}.

\begin{table}[ht]
\centering
\caption{\textbf{Data Statistics and Evaluation Protocols per Benchmark Task.} We detail the number of participants ($N$), total samples, and the specific validation strategies used for each task. Note that $N$ varies due to longitudinal attrition ($N=74$ at pre-intervention vs. $N=44$ at post-intervention).}
\label{tab:task_splits}
\small
\begin{tabular}{@{}lccccc@{}}
\toprule
\textbf{Task \& Description} & \textbf{N} & \textbf{Total Time} & \textbf{Intra} & \textbf{Inter} & \textbf{LOSO} \\ \midrule
\textbf{Task 1}: Cognitive State Decoding (Rest vs. Med.) & 74 & 14h 48mins$^\dagger$ & \checkmark & \checkmark & \checkmark \\
\textbf{Task 2 (Pre)}: Technique Classification & 74 & 9h 52mins$^\dagger$ & -- & \checkmark & -- \\
\textbf{Task 2 (Post)}: Technique Classification & 44 & 5h 52mins$^\dagger$ & -- & \checkmark & -- \\
\textbf{Task 3}: Cross-Session Adaptation & 44 & 8h 48mins$^\dagger$ & \checkmark & -- & -- \\ \bottomrule
\multicolumn{6}{l}{\scriptsize $^\dagger$ Approximation total duration}
\end{tabular}
\end{table}

\subsection{Task 1: Cognitive State Decoding (Rest vs. Focused Attention)}
This task evaluates if EEG features can robustly distinguish between closed-eye resting (restCE01) and focused attention meditation (slMedita). Closed-eye resting serves as a proxy for mind wandering, as participants were explicitly instructed to allow their thoughts to wander freely during this phase (Appendix~\ref{EEG session detail}). This task assesses model capacity to differentiate between undirected mind wandering and directed cognitive focus. We utilize data exclusively from the initial pre-intervention session to isolate acute physiological state differences and prevent confounding from long-term neuroplastic changes. This selection maximizes inter-subject diversity by including the full cohort before post-intervention attrition. Furthermore, the volume of continuous recordings provides sufficient data density for model convergence. Classification is evaluated using intra-subject, inter-subject, and leave-one-subject-out frameworks. We compare a global model against technique-specific models for HK, SA, and BF subgroups to determine if isolating specific meditation techniques effectively reduces cross-subject variance (Appendix \ref{benchmark_detail}).

\subsection{Task 2: Fine-Grained Technique Classification and Longitudinal Tracking}

This task classifies specific meditation techniques: HK, SA, or BF, based on EEG recordings captured during practice. Intra-subject and leave-one-subject-out (LOSO) evaluations are logically excluded because each participant practices a single technique (Appendix~\ref{benchmark_detail}). Since individual data lack multi-class labels, and LOSO test sets would render standard performance metrics undefined, we utilize an inter-subject framework via 5-fold cross-validation. To investigate longitudinal effects, we perform this classification (training/testing) separately from pre-intervention recordings ($N=74$) and post-intervention follow-ups ($N=44$). Comparing performance across these temporal milestones quantifies whether long-term training induces a divergence or convergence in the neural signatures associated with different techniques. 

\subsection{Task 3: Cross-Session Generalization and Domain Adaptation}
\label{task3_label}
This task evaluates model robustness across a six-week longitudinal gap, addressing signal non-stationarity and temporal concept drift. We utilize Task 1 intra-subject models, trained on pre-intervention data, within a transfer learning framework. To evaluate temporal drift independently of inter-subject variability, the evaluation is restricted to an intra-subject paradigm across two tiers:

\paragraph{Zero-Shot Generalization.} Pre-trained models from Task 1 are applied directly to unseen post-intervention data to establish a baseline for six-week temporal shift degradation.

\paragraph{Few-Shot Calibration.} Pre-trained models are fine-tuned using a minimal, class-balanced calibration subset of post-intervention data. Performance recovery is then evaluated on the remaining held-out data using two adaptation strategies: (1) full-model fine-tuning and (2) fine-tuning only the batch normalization (BN) layers and linear classification head. Details on calibration subset construction and fine-tuning implementation are provided in Appendix~\ref{benchmark_detail} and ~\ref{app:task3}.

This benchmark identifies the nature of longitudinal drift. Rapid recovery after calibration suggests fundamental stability of neural representations, attributing degradation to superficial domain shifts like electrode placement or impedance. Conversely, failure to recover indicates that the six-week intervention induced neuroplasticity, fundamentally altering functional neural representations and rendering historical decision boundaries obsolete.

\vspace{0.2in}
\section{Baseline Models}

To comprehensively benchmark the L-FAME dataset and provide a standard reference for the performance of the proposed tasks, we evaluate a diverse set of baseline models. These models are categorized into traditional machine learning approaches and end-to-end deep learning architectures.

\subsection{Classification Methods}

\paragraph{Traditional Machine Learning}
For the classification of mental states and meditation techniques, frequency-domain features typically serve as the most robust biomarkers. Therefore, our traditional baseline employs Power Spectral Density (PSD) features extracted across standard EEG frequency bands, such as theta, alpha, beta, and gamma. These handcrafted features are subsequently flattened and fed into standard classifiers, specifically Support Vector Machines (SVM). While we also evaluated the Filter Bank Common Spatial Pattern (FBCSP) coupled with an SVM as a spatial-domain reference, we use the features extracted from this method to train the SVMs following the intra-subject and inter-subject evaluations.

\paragraph{Deep Learning}
To evaluate automatic feature extraction without prior manual feature engineering, we benchmark four representative deep learning architectures, ranging from standard convolutional networks to advanced spatiotemporal models. First, we utilize Shallow ConvNet and Deep ConvNet, two classical architectures widely adopted in BCI research~\cite{schirrmeister2017deep}. Shallow ConvNet extracts features analogous to band power, whereas Deep ConvNet utilizes a deeper hierarchy of generic convolutional layers to capture complex representations. Second, we deploy EEGNet, a compact and specialized convolutional neural network tailored for EEG signals~\cite{lawhern2018eegnet}. It utilizes depthwise and separable convolutions to significantly reduce trainable parameters while maintaining robust cross-subject generalization. Third, we implement EEG-Conformer, a state-of-the-art architecture integrating the local feature extraction of convolutions with the global, long-range dependency modeling of Transformer self-attention mechanisms~\cite{song2022eeg}. This provides an advanced baseline for complex spatiotemporal EEG decoding.

%%%%%%%%%%%%%%%%%%%%%%%%%%%%%%%%%%%%%%%%%%%%%%%%%%%%%%%%%%%%

\section{Results}

% \begin{table}[ht]
% \centering
% \caption{\small \textbf{Performance Comparison for Task 1: State Decoding.} We report the mean AUC (\%) and standard deviation across three evaluation protocols: Intra-subject, Inter-subject, and Leave-One-Subject-Out (LOSO) ($N=74$). The theoretical chance level for this binary classification task is 50.0\%.}
% \label{tab:task1_results}
% \small
% \renewcommand{\arraystretch}{1.2}
% \setlength{\tabcolsep}{8pt}
% \begin{tabular}{@{} l ccc @{}}
% \toprule
% \textbf{Model} & \textbf{Intra-subject} & \textbf{Inter-subject} & \textbf{LOSO} \\ 
% \midrule
% PSD + SVM / RF    & -- & -- & -- \\ 
% FBCSP + SVM / RF  & -- & -- & -- \\ 
% \midrule
% ShallowConvNet    & 97.0{\scriptsize $\pm$2.0} & 66.5{\scriptsize $\pm$4.3} & 70.4{\scriptsize $\pm$20.9} \\
% DeepConvNet       & 97.0{\scriptsize $\pm$1.5} & 66.2{\scriptsize $\pm$5.9} & 68.4{\scriptsize $\pm$27.0} \\
% EEGNet            & 99.2{\scriptsize $\pm$1.2} & 66.5{\scriptsize $\pm$4.0} & 66.6{\scriptsize $\pm$27.4} \\
% EEG-Conformer     & 97.2{\scriptsize $\pm$1.4} & 66.9{\scriptsize $\pm$5.2} & 67.5{\scriptsize $\pm$24.5} \\ 
% \bottomrule
% \end{tabular}
% \end{table}

\begin{wraptable}{r}{0.55\textwidth} 
    \centering
    \vspace{-12pt} 
    \caption{\small \textbf{Task 1: State Decoding Performance.} Mean AUC (\%) and standard deviation for Intra-subject, Inter-subject, and LOSO protocols ($N=74$). }
    \label{tab:task1}
    \small
    \renewcommand{\arraystretch}{1.2}
    \setlength{\tabcolsep}{3pt} 
    \begin{tabular}{@{} l ccc @{}}
        \toprule
        \textbf{Model} & \textbf{Intra} & \textbf{Inter} & \textbf{LOSO} \\ 
        \midrule
        PSD + SVM    & 97.1{\scriptsize $\pm$4.0}  & 59.4{\scriptsize $\pm$3.8}   & 61.3{\scriptsize $\pm$4.4}  \\
        FBCSP + SVM  & \textbf{99.9{\scriptsize $\pm$0.2}} & 55.8{\scriptsize $\pm$6.2} & 58.2{\scriptsize $\pm$21.4} \\ 
        \midrule
        ShallowConvNet  & 97.2{\scriptsize $\pm$4.4} & 66.5{\scriptsize $\pm$3.9} & \textbf{70.4{\scriptsize $\pm$20.8}} \\
        DeepConvNet     & 97.0{\scriptsize $\pm$8.5} & 66.2{\scriptsize $\pm$5.3} & 68.4{\scriptsize $\pm$26.8} \\
        EEGNet          & 99.2{\scriptsize $\pm$2.1} & 66.5{\scriptsize $\pm$3.5} & 66.6{\scriptsize $\pm$27.2} \\
        EEG-Conformer   & 97.0{\scriptsize $\pm$4.9} & \textbf{66.9{\scriptsize $\pm$4.7}} & 67.5{\scriptsize $\pm$24.3} \\ 
        \bottomrule
    \end{tabular}
    \vspace{-10pt} 
\end{wraptable}

  % PSD + SVM / RF    & -- & -- & -- \\ 

\paragraph{Task 1: Cognitive State Decoding.} Table \ref{tab:task1} presents the fundamental capacity of various models to distinguish between the resting state and the focused attention meditation state. Deep learning architectures demonstrate exceptional proficiency in the intra-subject evaluation paradigm, with EEGNet achieving a peak AUC of $99.2\% \pm 2.1\%$, closely followed by ShallowConvNet at $97.2\% \pm 4.4\%$. However, a substantial generalization gap emerges when evaluating models across subjects. Inter-subject and LOSO evaluations yield significantly lower performance, hovering around $66\%$ to $70\%$. This pronounced variance underscores the highly individualized nature of meditation-induced neural oscillations. Furthermore, traditional machine learning baselines relying on handcrafted features, such as PSD and FBCSP coupled with SVM, exhibit optimal performance compared to deep models in the intra-subject evaluation case (yielding an estimated AUC of $\approx 99.9\%$). However, they under-perform in inter-subject and LOSO evaluations, rendering them inferior to convolutional networks in generalizing to unseen subjects. As detailed in Appendix~\ref{app:task1}, group-specific decoding reveals that the SA TA NA MA (SA) technique provides the most generalized neural signatures, yielding the highest LOSO accuracy ($69.1\% \pm 23.1\%$ via EEGNet) among the three practices.
\begin{wraptable}{r}{0.43\textwidth}
    \centering
    \vspace{-10pt}
    \captionsetup{width=0.\linewidth}
    \caption{\small \textbf{Task 2: Technique Classification Performance.} PR-AUC (\%) for Inter-subject evaluation (Pre: $N=74$, Post: $N=44$).}
    \label{tab:task2}
    \small
    \renewcommand{\arraystretch}{1.2}
    \setlength{\tabcolsep}{4pt} 
    \begin{tabular}{@{} l cc @{}}
        \toprule
        \textbf{Model} & \textbf{Pre} & \textbf{Post}\\
        \midrule

        FBCSP + SVM  & 34.3{\scriptsize $\pm$4.1} & 47.4{\scriptsize $\pm$7.2} \\ 
        \midrule
        ShallowConvNet  & \textbf{38.0{\scriptsize $\pm$6.2}} & 45.7{\scriptsize $\pm$9.7} \\
        DeepConvNet     & 34.5{\scriptsize $\pm$3.4} & 45.8{\scriptsize $\pm$10.7} \\
        EEGNet          & 35.0{\scriptsize $\pm$2.9} & \textbf{48.8{\scriptsize $\pm$8.6}} \\
        EEG-Conformer   & 37.6{\scriptsize $\pm$4.6} & 43.9{\scriptsize $\pm$11.4} \\ 
        \bottomrule
    \end{tabular}
    \vspace{-10pt} 
\end{wraptable}

\paragraph{Task 2: Fine-Grained Technique Classification}Classifying the specific meditation technique directly from the EEG signals represents a significantly more complex challenge. As shown in Table \ref{tab:task2}, all models yield Precision-Recall AUC (PR-AUC) scores that, while exceeding the theoretical chance level of $33.3\%$, remain generally low. In the pre-intervention phase, ShallowConvNet achieves the highest baseline performance at $38.0\% \pm 6.2\%$. Interestingly, the post-intervention data ($N=44$) exhibits an observable shift in classification dynamics, with EEGNet improving to $48.8\% \pm 8.6\%$. This subtle enhancement suggests that the six-week longitudinal training intervention may gradually consolidate technique-specific neural signatures, albeit the representations remain inherently difficult to isolate across different individuals. We also notice that when fitting an SVM to the features extracted using FBCSP we also see an improvement in the PR-AUC when testing the post-intervention data. In the paired analysis restricted to the same 44 participants, EEGNet PR-AUC increases from $42.9\% \pm 8.7\%$ to $48.8\% \pm 8.6\%$, whereas ShallowConvNet shows no corresponding increase (Appendix~\ref{app:task2}). The improvement in EEGNet performance within the paired cohort is consistent with the possibility that technique-related EEG patterns become more distinguishable following six weeks of training. The absence of a corresponding increase in ShallowConvNet suggests that sensitivity to these pre-to-post differences may depend on the model architecture and its learned representations.

% \begin{wraptable}[12]{r}{0.65\textwidth}
% \centering
% \small
% \renewcommand{\arraystretch}{0.9}
% \setlength{\tabcolsep}{4pt}
    
% \caption{\small \textbf{Task 3: Cross-Session Generalization and Longitudinal Adaptation.} Note that the $k$-shot calibration performance reported in this table corresponds to the full fine-tuning strategy (updating all layers). }
% \label{tab:task3_fewshot}
% \small
% \begin{tabular}{@{}lcccc@{}}
% \toprule
%  & \multicolumn{4}{c}{\textbf{Task 3: Intra-subject Adaptation (AUC \%)}} \\ 
%  \cmidrule(lr){2-5}
% \textbf{Model} & Zero-shot & 10-shot & 30-shot & Upper Bound$^\dagger$\\ \midrule
% ShallowConvNet & 63.9{\scriptsize $\pm$16.8} & 75.2{\scriptsize $\pm$17.5} &  78.3{\scriptsize $\pm$16.6} & 95.4{\scriptsize $\pm$7.2}\\
% DeepConvNet    & 63.1{\scriptsize $\pm$22.7} & 72.8{\scriptsize $\pm$20.2} & 77.9{\scriptsize $\pm$18.4}& 97.0{\scriptsize $\pm$5.8} \\
%  EEGNet         & 63.1{\scriptsize $\pm$17.8} & \textbf{76.5{\scriptsize $\pm$21.2}} & \textbf{79.3{\scriptsize $\pm$21.6}} &95.4{\scriptsize $\pm$8.3} \\
% EEG-Conformer      & \textbf{67.2{\scriptsize $\pm$16.5}} & 74.5{\scriptsize $\pm$18.1}  & 77.2{\scriptsize $\pm$18.0} & 93.3{\scriptsize $\pm$7.9} \\ 
% \bottomrule
% \multicolumn{5}{l}{\scriptsize $^\dagger$ representing the performance ceiling achieved through full intra-blockwise training.}

% \end{tabular}
% \end{wraptable}

\vspace{0.1in}

\paragraph{Task 3: Cross-Session Generalization and Longitudinal Adaptation}
Table~\ref{tab:task3_fewshot} summarizes cross-session state-decoding performance following the six-week intervention. Under zero-shot evaluation, models achieve AUCs between $63.1\%$ and $67.2\%$, substantially below their within pre-intervention performance. This decline indicates that models trained on pre-intervention data do not generalize well to post-intervention recordings. To address this gap, we compared full fine-tuning, which updates all model parameters, with BN-and-head fine-tuning, which updates the BatchNorm affine parameters and the linear classification head. Table~\ref{tab:task3_fewshot} reports the full fine-tuning results, while a detailed comparison between the two strategies is provided in Appendix~\ref{app:task3}. With $30$ calibration windows per class, full fine-tuning improves EEGNet AUC from $63.1\%$ to $79.3\%$. However, performance across the four architectures remains $16.1$ to $19.1$ percentage points below the within pre-intervention performance. These results show that limited post-intervention calibration improves cross-session decoding, although it does not fully recover within-session performance.

\begin{table}[H]
    \centering
    \small
    \renewcommand{\arraystretch}{1.1}
    \setlength{\tabcolsep}{10pt}

    \caption{\textbf{Task 3: Cross-session generalization and adaptation.} AUC (\%) under zero-shot evaluation and full fine-tuning with 10 or 30 calibration windows per class.}
    \label{tab:task3_fewshot}

    \begin{tabular}{@{}lcccc@{}}
        \toprule
        \textbf{Model} & \textbf{Zero-shot} & \textbf{10-shot} & \textbf{30-shot} & \textbf{Upper Bound}$^\dagger$ \\
        \midrule
        ShallowConvNet & $63.9 \pm 16.8$ & $75.2 \pm 17.5$ & $78.3 \pm 16.6$ & $95.4 \pm 7.2$ \\
        DeepConvNet & $63.1 \pm 22.7$ & $72.8 \pm 20.2$ & $77.9 \pm 18.4$ & $97.0 \pm 5.8$ \\
        EEGNet & $63.1 \pm 17.8$ & $\mathbf{76.5 \pm 21.2}$ & $\mathbf{79.3 \pm 21.6}$ & $95.4 \pm 8.3$ \\
        EEG-Conformer & $\mathbf{67.2 \pm 16.5}$ & $74.5 \pm 18.1$ & $77.2 \pm 18.0$ & $93.3 \pm 7.9$ \\
        \bottomrule
    \end{tabular}

    \par\smallskip
    {\footnotesize $^\dagger$Within-session reference performance obtained through full intra-blockwise training.\par}
\end{table}

%%%%%%%%%%%%%%%%%%%%%%%%%%%%%%%%%%%%%%%%%%%%%%%%%%%%%%%%%%%%
\section{Discussion and Future Work}

The L-FAME benchmark provides a standardized framework to evaluate EEG-based models across three progressively challenging tasks: the decoding of cognitive states (Task 1), fine-grained technique classification (Task 2), and cross-session generalization (Task 3). Across all tasks, a consistent pattern emerges: current models achieve high intra-subject performance but degrade significantly when generalizing across subjects or time. Although Task 1 intra-subject AUC exceeds 97\%, cross-subject performance plateaus between 66\% and 70\%. This gap of approximately 30\% reveals a fundamental discrepancy between standard laboratory evaluations and real-world deployment requirements. Furthermore, Task 2 remains an unresolved cross-subject challenge even after the intervention. Task 3 indicates that longitudinal drift is substantial but remains largely correctable through minimal calibration. Appendices~\ref{app:task1} to \ref{app:task3} provide detailed subgroup analyses, representational geometry experiments, and adaptation ablations for each task.

These results align with the established perspective that meditation-induced EEG states exhibit strong within-individual signatures~\cite{shang2023eeg,rodriguez2024assessing}. They also confirm that inter-individual neural variability~\cite{huang2023discrepancy,haegens2014inter} and temporal non-stationarity~\cite{krusienski2011critical} remain primary bottlenecks. The difficulty of cross-subject technique classification appears to conflict with prior studies reporting technique-specific EEG signatures~\cite{li2026not,ventura2024intrinsic,rodriguez2021eeg}. However, those studies primarily established group-level effects by aggregating within-subject relative changes, rather than evaluating generalized decision boundaries via end-to-end cross-subject classification. This distinction highlights a critical gap between statistical neurophysiological characterization and deploying robust decoding models. Notably, post-intervention representational geometry analyses (UMAP, RSA) demonstrate that six weeks of training initiates the differentiation of technique-specific neural representations. Specifically, the similarity margin between within-class and between-class representations expands from 0.30 to 1.10 (Appendix~\ref{app:task2}, Figure~\ref{fig:umap}). This expansion suggests that while these signatures are emerging, they remain unstable across subjects. An informative asymmetry also emerges: Breath Focus, which relies on somatic attention, generates more separable cross-subject representations than the two mantra-based techniques. 
%In contrast, HK and SA representations remain mutually confusable after the intervention. 
Although HK and SA representations become more distinct after the intervention, with their centroid cosine similarity decreasing from 0.953 to 0.250 (Appendix~\ref{app:task2}, Figure~\ref{fig:rsa}), distinguishing the two mantra-based techniques remains challenging in cross-subject classification, as supported by the near-chance one-vs-all performance (Appendix~\ref{app:task2}, Figure~\ref{fig:task2_ova_group}). This observation implies that the underlying cognitive substrate (somatic versus phonological) is potentially more determinative of cross-subject decoding performance than the specific content of the mantra~\cite{logie2003brain,lutz2008attention}.

The L-FAME benchmark highlights three concrete open problems for the research community. First, the persistent performance gap between intra-subject and cross-subject evaluations motivates the development of domain generalization and transfer learning methods. These methods must be tailored to spontaneous, internally driven cognitive states, which are poorly supported by existing benchmarks based on motor imagery or stimulus-evoked paradigms. Second, the near-chance performance in the cross-subject classification of techniques indicates that general-purpose architectures lack the inductive biases required to capture technique signatures that remain invariant across individuals. The integration of meditation/cognition-specific priors may be necessary. Third, Task 3 demonstrates that models can transfer over time despite the expected longitudinal EEG drift. This drift can be corrected using 30 calibration windows per class (60 windows in total) through full fine-tuning (Appendix~\ref{app:task3}), which presents a practical pathway toward the efficient deployment of longitudinal brain-computer interfaces without complete model retraining. More broadly, the L-FAME benchmark serves as a standardized resource for the investigation of training-induced neuroplasticity and the personalized monitoring of meditation. Both directions hold increasing relevance for the field of computational mental health~\cite{liu2025leveraging,lyu2026deep}.

%%%%%%%%%%%%%%%%%%%%%%%%%%%%%%%%%%%%%%%%%%%%%%%%%%%%%%%%%%%%
\section{Conclusion and Limitations}
We introduced L-FAME, a novel 6-week longitudinal EEG dataset and benchmark designed to evaluate neural decoding across three focused attention meditation practices. Our extensive baselines reveal a critical insight: while deep learning models achieve exceptional intra-subject accuracy, they struggle significantly with cross-subject generalization due to inherent inter-individual variability. Furthermore, we demonstrated that longitudinal temporal drift is substantial but can be effectively mitigated via minimal few-shot calibration. 

\textbf{Limitations \& Future Work:} Although attrition reduced the longitudinal cohort to 44 participants, our analyses found no statistically significant associations between dropout and the assessed baseline demographic or EEG features, providing no statistically significant evidence of selective attrition (see Appendix~\ref{MCAR}). Nonetheless, reliance on self-reported practice adherence, sparse temporal sampling (two discrete time points), the lack of a control group, and a restricted college-aged demographic limit the broader clinical generalizability of the findings. We plan to continue data collection in future work to supplement the dataset with appropriate control cohorts. We anticipate that this open-access resource will facilitate the development of meditation-specific architectures and robust domain-adaptation methods for continuous neurophysiological monitoring.

\section*{Acknowledgments}
We would like to acknowledge Devin O'Rourke and Sidharth Chhabra from The Harmony Collective, Ypsilanti, Michigan for their expert guidance in meditation training. We also extend our gratitude to Michigan State University students Ab Basit Rafi Syed, Pratham Pradhan, Annie Wozniak, Vu Song Thuy Nguyen, Genevieve Orlewicz, and Alisia Coipel for their valuable assistance with data collection.

\section*{Ethics statement}
All experimental procedures were approved by the Institutional Review Board (IRB) of Michigan State University. All participants provided written informed consent prior to the commencement of the study. The privacy rights of all human subjects have been observed throughout the research.
%%%%%%%%%%%%%%%%%%%%%%%%%%%%%%%%%%%%%%%%%%%%%%%%%%%%%%%%%%%%
% \section*{References}
\newpage
\bibliographystyle{unsrt}
\bibliography{references}

%%%%%%%%%%%%%%%%%%%%%%%%%%%%%%%%%%%%%%%%%%%%%%%%%%%%%%%%%%%%
\newpage
\appendix
\section*{Appendix}
This appendix provides supplementary material organized into four sections. 
The first section presents comprehensive dataset documentation adhering to the Datasheets for Datasets framework, which encompasses detailed participant demographics (Section~\ref{Detailed_Participant}), attrition analysis with MCAR validation (Section~\ref{MCAR}), and the complete BIDS-compliant dataset structure alongside its processing derivatives (Section~\ref{structure dataset}). 
The second section details the extended experimental paradigms, including criteria for participant inclusion and exclusion (Section~\ref{Criteria}), three meditation techniques and the corresponding training protocols (Section~\ref{techniques}), procedures for EEG sessions (Section~\ref{EEG session detail}), and outcomes of the psychometric questionnaires (Section~\ref{questionnaires}). 
The third section outlines the methodology, data quality check (QC) (Section~\ref{app:eeg-qc}) and the data processing pipeline in detail, encompassing parameters for EEG preprocessing (Section~\ref{data_preprocess}), configurations for model training (Section~\ref{training_detail}), protocols for benchmark tasks (Section~\ref{benchmark_detail}), and details regarding hardware and software implementations. 
The fourth section reports additional experimental results for all three benchmark tasks, featuring ablations of the evaluation protocols (Section~\ref{app:task1}), longitudinal representational analyses (Section~\ref{app:task2}), and learning curves for few-shot adaptation (Section~\ref{app:task3}).

\section{Dataset Documentation (Datasheets for Datasets)}

\subsection{Detailed Participant Demographics}
\label{Detailed_Participant}
\begin{figure}
    \centering
    \includegraphics[width=1\linewidth]{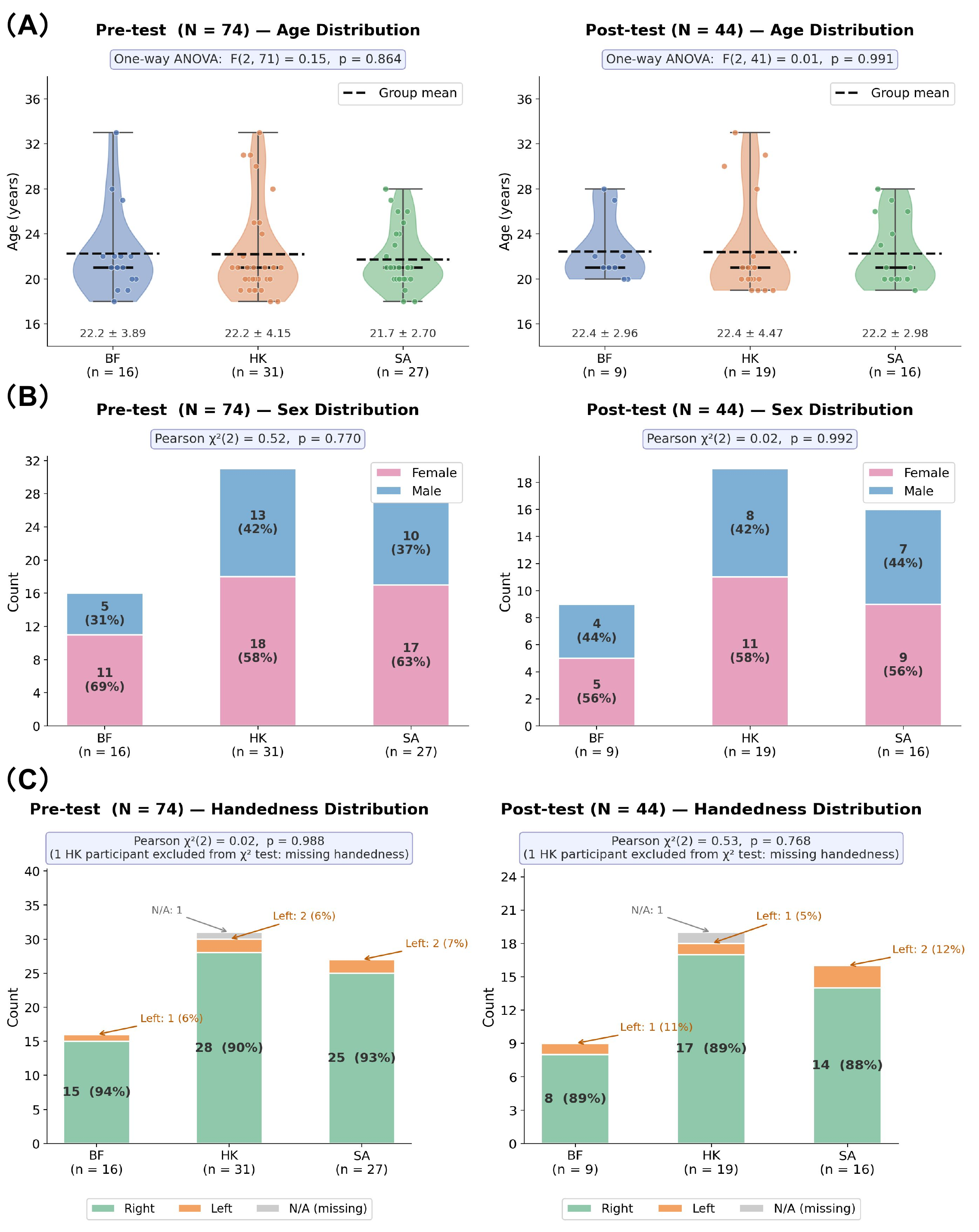}
    \caption{Distributions of (A) age, (B) sex, and (C) handedness demonstrate consistent demographic homogeneity across intervention groups at both pre-test and post-test phases.}
    \label{fig:age_sex_hand}
\end{figure}

\paragraph{Baseline Demographic Homogeneity} A total of 74 participants meeting the inclusion criteria (no history of neurological or psychiatric disorders, and no or minimal prior meditation experience) were initially recruited and randomly assigned to either the Breath Focus meditation (BF), Hare Krishna mantra meditation (HK) or SA-TA-NA-MA mantra meditation (SA) group (see in Table~\ref{tab:detail participants}). Statistical analyses via one-way ANOVA and Pearson $\chi^2$ tests confirm that the three groups were no statistically significant baseline differences were detected at the pre-test phase. Specifically, no significant differences were observed across groups regarding age ($F(2,71)=0.15$, $p=0.864$), sex distribution ($\chi^2(2)=0.52$, $p=0.770$), or handedness ($\chi^2(2)=0.02$, $p=0.988$, with one HK participant excluded from this specific metric due to missing data) in Figure~\ref{fig:age_sex_hand} first column. The overall cohort primarily consisted of right-handed young adults (mean age roughly 22 years) with a slightly higher proportion of females ($62.2\%$).

\paragraph{Consistent Attrition and Preserved Post-Test Balance} The study experienced a notable overall attrition rate of 40.5\% between the pre-test and post-test phases, reducing the final sample size to 44 participants. However, the dropout rates were distributed evenly across the intervention groups, ranging from 38.7\% in the HK group to 43.8\% in the BF group. Importantly, homogeneity tests on the post-test sample indicate that this attrition was non-selective and did not introduce demographic bias. The remaining participants across the three groups continued to show no significant differences in age ($F(2,41)=0.01$, $p=0.991$), sex ($\chi^2(2)=0.02$, $p=0.992$), and handedness ($\chi^2(2)=0.53$, $p=0.768$) in Figure~\ref{fig:age_sex_hand} second column. This preserved demographic parity across time points ensures that downstream physiological analyses (such as EEG signal evaluation) remain robust against confounding demographic variables, thereby supporting the interpretability and fairness of subsequent evaluations.
\begin{table}[ht]
    \centering
    \caption{Participant Demographics and Homogeneity Test Results by Group and Assessment Phase}
    \label{tab:demographics}
    \setlength{\tabcolsep}{8pt} % 稍微加宽间距以适应新的布局
    \begin{tabular}{@{}llcccc@{}}
        \toprule
        \multirow{2}{*}{Phase} & \multirow{2}{*}{Variable} 
        & \multicolumn{3}{c}{Group} 
        & \multirow{2}{*}{Total} \\
        \cmidrule(lr){3-5}
        & & BF & HK & SA & \\
        \midrule
        \multirow{8}{*}{\parbox{1.6cm}{\centering\textbf{Pre-test}\\$N=74$}}
        & $N$ & 16 & 31 & 27 & 74 \\
        & Age, M$\pm$SD 
            & $22.25\pm3.89$ 
            & $22.19\pm4.15$ 
            & $21.74\pm2.70$ 
            & $22.04\pm3.58$ \\[2pt]
        & \multicolumn{5}{l}{\textit{Sex}} \\
        & \quad Female, $n$\,(\%) 
            & 11 (68.8) 
            & 18 (58.1) 
            & 17 (63.0) 
            & 46 (62.2) \\
        & \quad Male, $n$\,(\%) 
            & 5 (31.2) 
            & 13 (41.9) 
            & 10 (37.0) 
            & 28 (37.8) \\[2pt]
        & \multicolumn{5}{l}{\textit{Handedness}} \\
        & \quad Right, $n$\,(\%) 
            & 15 (93.8) 
            & 28 (90.3) 
            & 25 (92.6) 
            & 68 (91.9) \\
        & \quad Left, $n$\,(\%) 
            & 1 (6.2) 
            & 2 (6.5) 
            & 2 (7.4) 
            & 5 (6.8) \\
        & Attrition, $n$\,(\%) 
            & 7 (43.8) 
            & 12 (38.7) 
            & 11 (40.7) 
            & 30 (40.5) \\
        \midrule
        \multirow{7}{*}{\parbox{1.6cm}{\centering\textbf{Post-test}\\$N=44$}}
        & $N$ & 9 & 19 & 16 & 44 \\
        & Age, M$\pm$SD 
            & $22.44\pm2.96$ 
            & $22.37\pm4.47$ 
            & $22.25\pm2.98$ 
            & $22.34\pm3.62$ \\[2pt]
        & \multicolumn{5}{l}{\textit{Sex}} \\
        & \quad Female, $n$\,(\%) 
            & 5 (55.6) 
            & 11 (57.9) 
            & 9 (56.2) 
            & 25 (56.8) \\
        & \quad Male, $n$\,(\%) 
            & 4 (44.4) 
            & 8 (42.1) 
            & 7 (43.8) 
            & 19 (43.2) \\[2pt]
        & \multicolumn{5}{l}{\textit{Handedness}} \\
        & \quad Right, $n$\,(\%) 
            & 8 (88.9) 
            & 17 (89.5) 
            & 14 (87.5) 
            & 39 (88.6) \\
        & \quad Left, $n$\,(\%) 
            & 1 (11.1) 
            & 1 (5.3) 
            & 2 (12.5) 
            & 4 (9.1) \\
        \bottomrule
    \end{tabular}
    \label{tab:detail participants}
\end{table}

\subsection{Attrition Cohorts Analysis and MCAR Validation}
\label{MCAR}

\begin{figure}[ht]
    \centering
    \includegraphics[width=\linewidth]{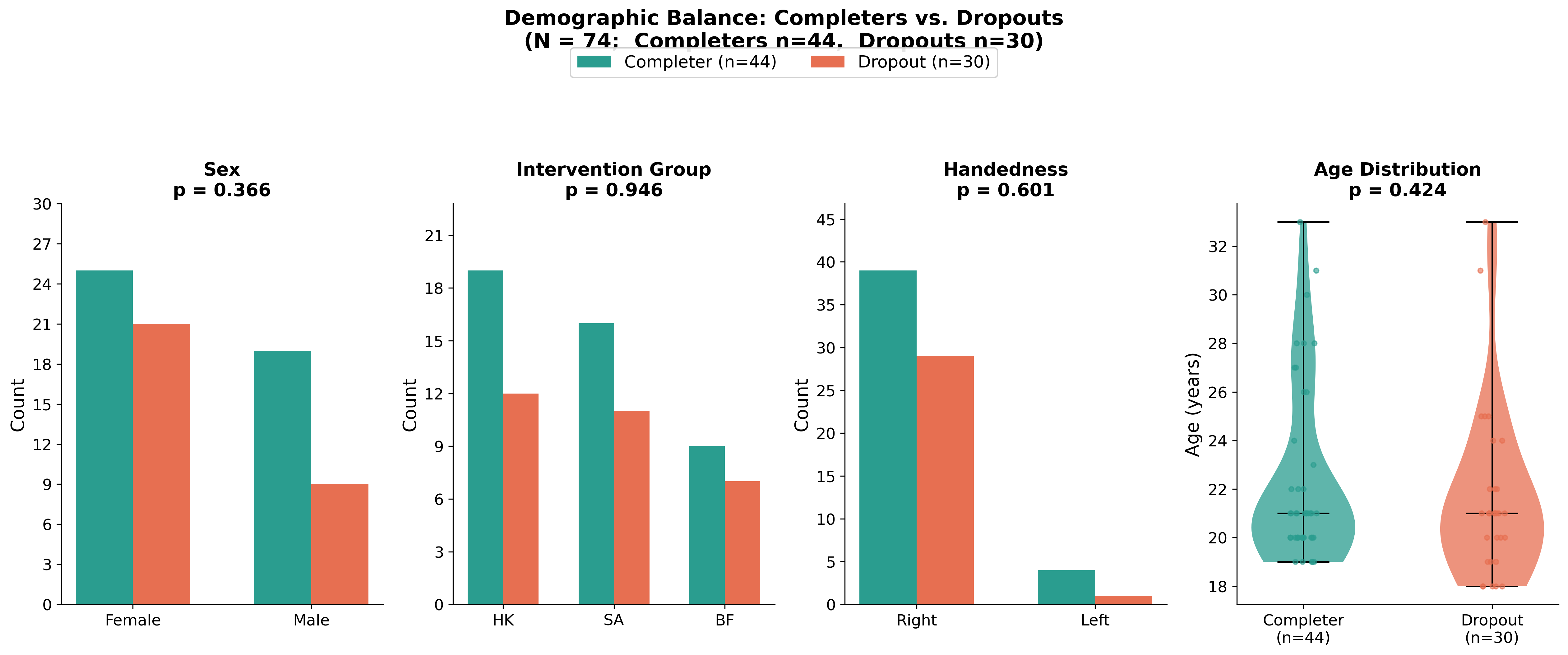}
    \caption{\small Analysis of Group Assignment and Demographics as Potential Causes of Dropout and Attrition, Detailing the Distributions of Sex, Group, Handedness, and Age With Corresponding $p$-Values for Each Analysis}
    \label{fig:demographic_attr}
\end{figure}

\paragraph{Baseline Demographics and Group Assignments Exhibit No Significant Correlation with Attrition}
To determine whether participant attrition was biased by baseline characteristics, we compared the demographic profiles of completers ($n = 44$) and dropouts ($n = 30$) across four key dimensions (Fig.~\ref{fig:demographic_attr}). Pearson's $\chi^{2}$ tests of independence were conducted for categorical variables, revealing no significant associations between dropout status and sex ($p = 0.366$), intervention group ($p = 0.946$), or handedness ($p = 0.601$). For the continuous variable of age, a Mann-Whitney U test indicated no significant difference in the age distribution between the two cohorts ($p = 0.424$). These results demonstrate robust baseline parity, confirming that neither biological factors nor the specific nature of the assigned meditation protocol (HK, SA, or BF) served as a primary catalyst for dropout. The absence of demographic or group-level divergence supports the assumption that the observed attrition is likely missing completely at random (MCAR) relative to these factors, validating the subsequent focus on intrinsic neurophysiological markers for predicting participant adherence.

\begin{wrapfigure}[10]{r}{0.5\textwidth}
    \centering
    \includegraphics[width=0.75\linewidth]{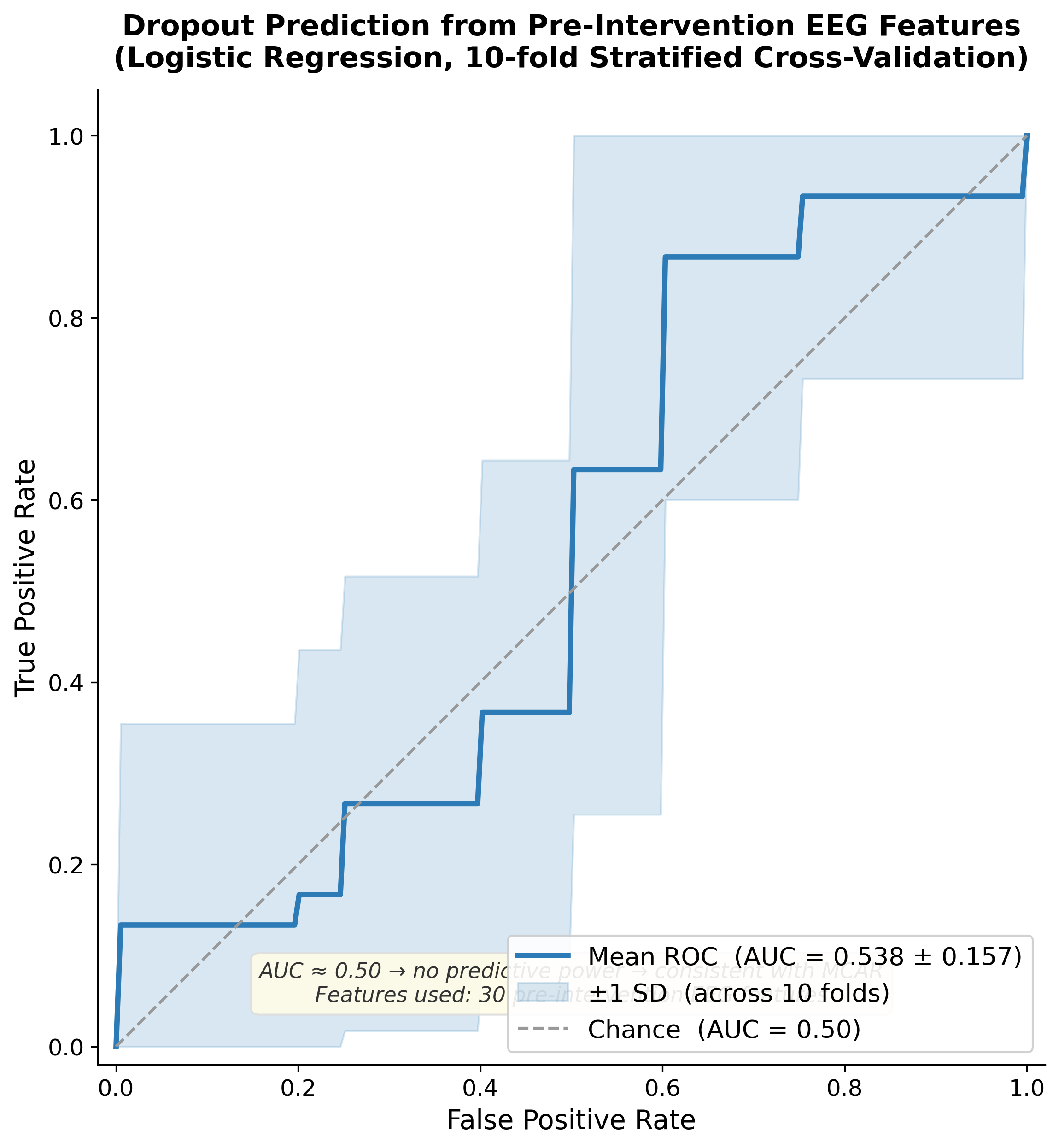}
    \caption{\small Receiver Operating Characteristic (ROC) curve for dropout prediction using pre-intervention EEG features.}
    \label{fig:prediction_drop}
\end{wrapfigure}

\begin{figure}[ht]
    \centering
    \includegraphics[width=\linewidth]{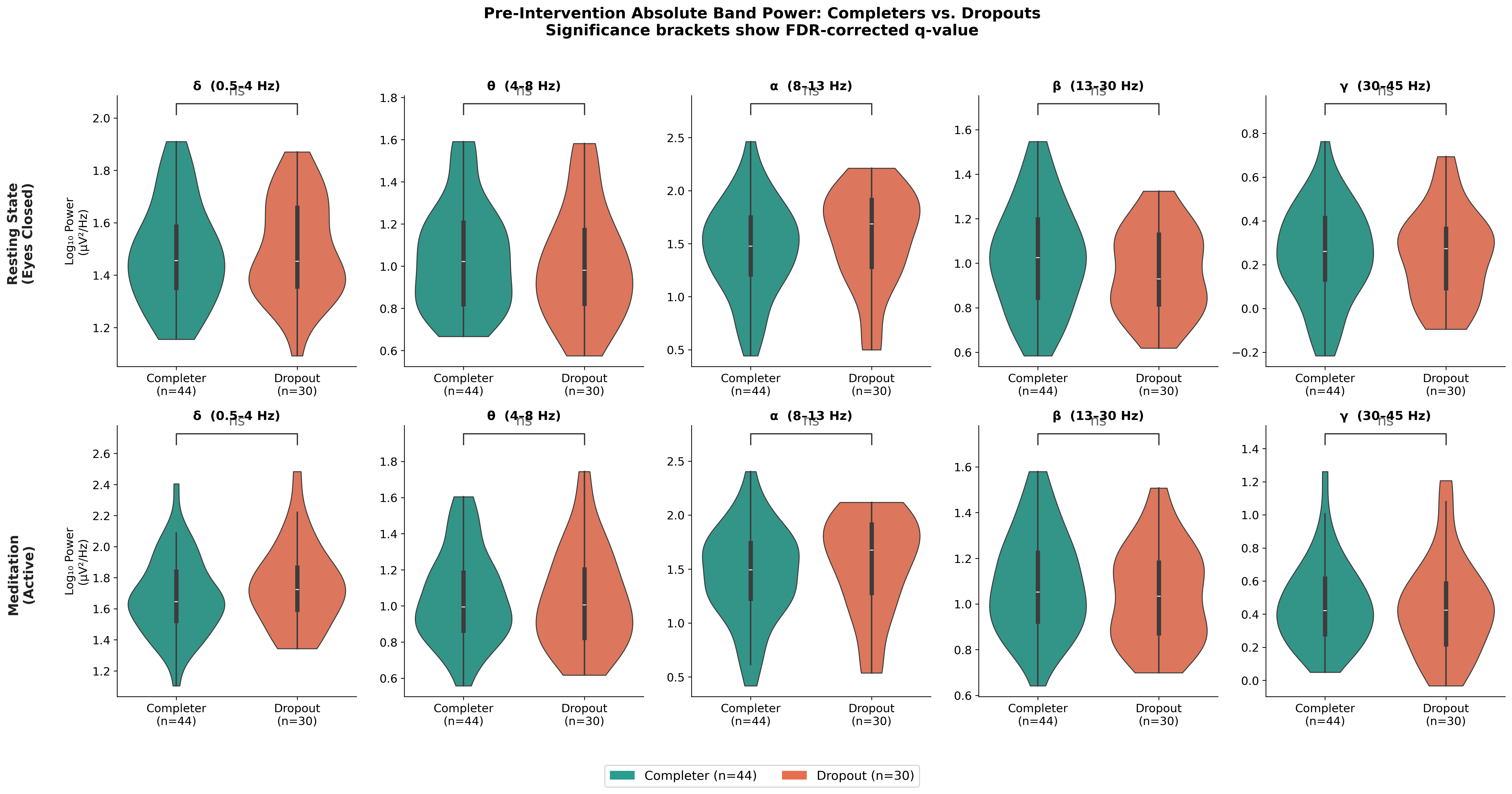}
    \caption{\small \textbf{Pre-intervention absolute EEG band power distributions for completers vs. dropouts.} Comparison of $\log_{10}$-transformed absolute power spectral density (PSD) between participants who completed the study ($n = 44$, teal) and those who dropped out ($n = 30$, orange). Results are shown for five canonical frequency bands under both Resting State (top row) and Active Meditation (bottom row) conditions. Significance brackets denote FDR-corrected $q$-values; "ns" indicates no statistically significant differences ($q > 0.05$) }
    \label{fig:band power}
\end{figure}

\paragraph{Pre-Intervention EEG Features Show Limited Predictive Value for Participant Attrition} To rigorously evaluate whether baseline neurophysiological states bias experimental retention, we analyzed 30 pre-intervention EEG features across both univariate and multivariate dimensions (Fig.~\ref{fig:band power}, Fig.~\ref{fig:prediction_drop}). Univariate comparisons of absolute power spectral density (PSD) across five canonical frequency bands ($\delta, \theta, \alpha, \beta, \gamma$) under resting-state and active meditation conditions revealed no significant differences between completers ($n=44$) and dropouts ($n=30$) after FDR correction (all $q > 0.05$, Fig.~3). Similarly, frontal $\theta$ power and Phase Locking Value (PLV) for key electrode pairs (F3--F4, Fz--Pz, mean frontal) exhibited functional homogeneity between cohorts~\cite{cavanagh2014frontal,lomas2015systematic,lachaux1999measuring}. To assess the collective discriminative power of these features, a logistic regression classifier with $\ell_{2}$ regularization was evaluated via 10-fold stratified cross-validation. The resulting mean AUC was $0.538 \pm 0.157$, which departs from the 0.50 chance baseline by only 0.038 units (Fig.~\ref{fig:prediction_drop}). 

The combination of non-significant univariate band-level differences and near-chance multivariate classification performance provides direct evidence against a Missing Not At Random (MNAR) mechanism. These results indicate that baseline neural representations carry effectively no predictive information regarding subsequent dropout, thereby validating the Missing Completely At Random (MCAR) assumption for the L-FAME dataset.

\vspace{1.5in}

\subsection{Dataset Structure and Processing Derivatives}
\label{structure dataset}
\begin{figure}[t]
    \centering
    \includegraphics[width=\linewidth]{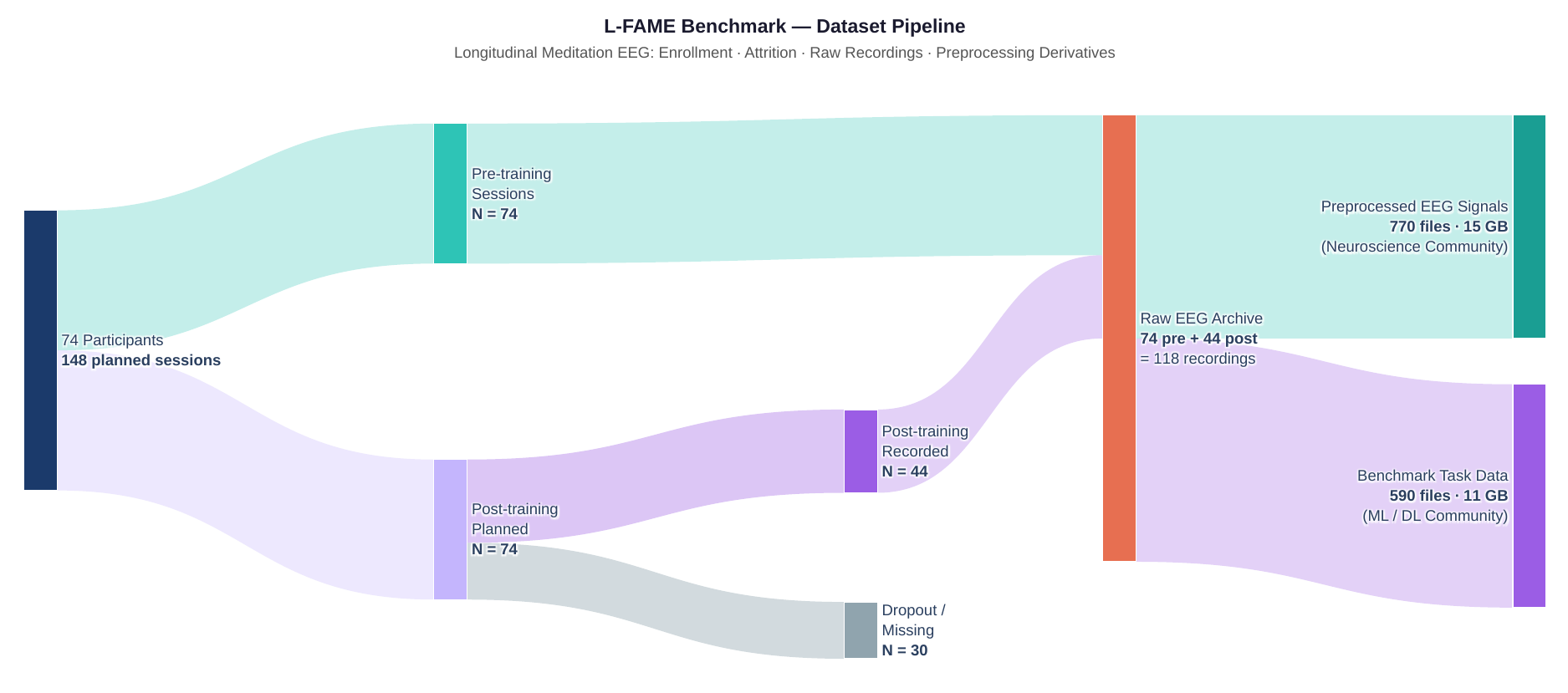}
    \caption{%
        Diagram of the Longitudinal Meditation Benchmark pipeline.
        Link widths are proportional to the recording counts at each stage.
        The diagram illustrates the progression from raw data acquisition,
        accounting for participant attrition, to the generation of the two
        preprocessed derivative tiers.
    }
    \label{fig:dataset-sankey}
\end{figure}

The Longitudinal Meditation Benchmark adheres to the Brain Imaging Data Structure (BIDS) v1.9.0 specification. It comprises recordings from 74 participants (\texttt{sub-01} through \texttt{sub-74}).

\paragraph{Root-level metadata.}
The root directory encapsulates standard BIDS metadata files, including \texttt{dataset\_description.json} for license and software provenance, \texttt{participants.tsv} and \texttt{participants.json} for per-subject demographics (such as group, age, sex, and handedness), \texttt{croissant\_metadata.json} serving as the ML dataset card, and a comprehensive \texttt{README} file.

\paragraph{Raw EEG recordings.}
The directory for each subject contains two longitudinal sessions: \texttt{ses-premedita} (pre-training) and \texttt{ses-posmedita} (post-training). As depicted in Figure~\ref{fig:dataset-sankey}, of the 148 planned sessions, all 74 pre-training sessions and 44 post-training sessions were successfully recorded. This attrition yields a total of 118 raw BrainVision \texttt{.eeg} files. Within each session, the EEG data are maintained in the BrainVision format, which consists of a binary data file (\texttt{.eeg}, approximately 114 to 125\,MB per session), a text header (\texttt{.vhdr}), and a marker file (\texttt{.vmrk}). These are accompanied by BIDS sidecar files detailing channel information, event metadata, electrode positions based on the CapTrak coordinate system, and acquisition parameters. Furthermore, four experimental conditions are recorded during each session: a sustained meditation task (\texttt{task-Medita}, \texttt{task-slMedita}) alongside resting-state blocks with eyes open (\texttt{task-restOE}) and eyes closed (\texttt{task-restCE01}, \texttt{task-restCE02}).

\paragraph{Preprocessed derivatives.}
The \texttt{derivatives/} directory houses two distinct tiers of processed data. Both processing pipelines operate on all 118 sessions across the five task segments, initially producing 590 base files per tier (Figure~\ref{fig:dataset-sankey}). The first tier, \texttt{eeglab\_preproc/} (15\,GB), contains data preprocessed via EEGLAB in \texttt{.set} and \texttt{.fdt} file pairs. The file count in this tier is elevated because it retains an intermediate post-ICA artifact-removal stage (\texttt{\_preproc\_icrm}) for the three tasks that require ICA cleaning (\texttt{task-slMedita}, \texttt{task-restCE01}, \texttt{task-restCE02}). The second tier, \texttt{ml\_preproc\_data/} (11\,GB, 590 files), discards these intermediate pairs, storing only the final cleaned epoch tensor as a single NumPy (\texttt{.npy}) array per session per task.

%% ======================================================================= %%

\section{Extended Experimental Paradigms and Questionnaires}

\subsection{Participant Inclusion and Exclusion Criteria}
\label{Criteria}
To ensure sample homogeneity and data integrity, participants were screened based on several physiological, neurological, and logistical criteria. Inclusion required an intermediate level of English proficiency or higher and a current affiliation with Michigan State University (e.g., as a student, faculty, or staff member). Candidates were excluded if they reported significant vision, speech, or hearing impairments, or a history of significant head injuries or neurological disorders, including epilepsy, seizures, or stroke. Furthermore, participants were required to be free of any medications known to alter brain function and were mandated to abstain from alcohol and THC-containing substances for 24 hours preceding the experimental session. Additional screening factors included summer availability, and documented meditation experience. Specifically, screening ensured that participants had little to no prior meditation experience (defined as having practiced only once, or for a few days at least five years prior). Final eligibility was also contingent upon the ability to have reliable transportation to the testing facility. Participants received \$20 in compensation at each of three study checkpoints, for a maximum total compensation of \$60 upon completing all three checkpoints.

\subsection{Meditation Techniques and Training \& Practice Protocol}
\label{techniques}
The present study employed three distinct meditative practices to investigate the neurophysiological correlates across the participants: breath focus meditation, Hare Krishna mantra meditation, and SA-TA-NA-MA mantra meditation. The practice of breath focus meditation requires the sustained attention of the practitioner on the somatic sensations of respiration, functioning primarily as an exercise in cognitive control and the continuous monitoring of sensory input. By repeatedly redirecting the focus of the mind back to the breath upon the detection of mind-wandering episodes, this technique actively recruits the executive control networks of the brain. Both the Hare Krishna and SA-TA-NA-MA mantra meditations involve the continuous mental or vocal repetition of a specific sequence of syllables, which, in addition to executive control, engages the articulatory rehearsal component of working memory and modulates the default mode network through constant auditory and cognitive engagement.

Prior to the formal recording of the EEG data, all participants underwent a standardized training protocol to ensure the accurate execution of each technique. The duration of the training session was approximately twenty-five minutes, utilizing a standardized audio script rather than live expert guidance to maintain absolute consistency across all subjects. During this phase, the participants were seated comfortably in a sound-attenuated room to minimize environmental distractions and physiological artifacts. The instructions detailed the required posture, the precise mechanics of the chants, and the standardized method for redirecting attention when distracted, thereby establishing a uniform experiential baseline for the participants before the commencement of the experimental trials. Following the initial baseline assessments, participants were instructed to practice daily following a gradually increasing schedule designed to enhance achievability: 5 minutes daily for the first week, 10 minutes daily for the second week, and 15 minutes daily from the third week onwards.

Furthermore, to monitor adherence and ensure practice quality throughout the intervention, participants maintained a daily online journal. This log systematically recorded the practice time of day, total duration, post-meditation thoughts or feelings, and a 1 to 5 self-rated assessment of overall focus quality.

\subsection{EEG Session and Task Procedures}
\label{EEG session detail}

During both the baseline and post-intervention EEG recording sessions, participants completed a standardized sequence of five tasks while seated comfortably. The total duration of the session was approximately one hour, encompassing the preparation and the recording. Prior to the commencement of each block, the research team provided oral instructions and manually inserted event markers into the continuous EEG recording to precisely denote the onset and offset of the tasks. To minimize oculomotor artifacts, participants were instructed to maintain eye closure throughout the sequence, with the exception of the initial baseline task.

\begin{enumerate}
    \item \textbf{Eyes-Open Resting State (restOE) - 2 mins:} For this initial baseline, participants were instructed to keep their eyes open and maintain a relaxed state.
    
    \item \textbf{Eyes-Closed Resting State 1 (restCE01) - 4 mins:} For this pre-task baseline, participants were instructed to \textit{``close your eyes and let your mind wander."}
    
    \item \textbf{Active Meditation (Medita) - 8 mins:} The instructions for this block were dependent on the assigned group. Participants in the mantra groups (SA and HK) were instructed to \textit{``close your eyes throughout the task, chant the assigned mantra out loud, and focus on the mantra."} Participants in the BF group were instructed to \textit{``close their eyes and perform alternate nostril breathing and focus on their breath"}.
    
    \item \textbf{Eyes-Closed Resting State 2 (restCE02) - 4 mins:} This post-meditation resting interval utilized instructions identical to those of restCE01.
    
    \item \textbf{Silent Meditation (slMedita) - 8 mins:} Participants in the SA and HK groups were instructed to \textit{``repeat the mantra in your mind, like inner speech, and focus on that with eyes closed throughout the task time."} Participants in the BF group were instructed to  \textit{``close your eyes and focus on your breathing"}. 
\end{enumerate}

\subsection{Psychometric Questionnaires}
\label{questionnaires}

To establish a comprehensive psychological profile and provide a standardized context for the interpreted neural activity, a battery of psychometric assessments was administered before the experimental trials. The first instrument administered was the Perceived Stress Scale (PSS) ~\cite{cohen1988perceived}. The primary function of this scale is to evaluate the degree to which situations in the daily life of the individual are appraised as stressful. By quantifying the unpredictable and uncontrollable aspects of the life of the respondent, the PSS provides a critical measure of the current psychological load, which serves as a potential covariate in the analysis of the efficacy of the meditation intervention.

Following the assessment of stress, the participants completed the Short Form of the Five Facet Mindfulness Questionnaire (FFMQ-SF)~\cite{bohlmeijer2011psychometric}. The purpose of this instrument is to characterize the inherent capacity of the individual for trait mindfulness. The study evaluates this disposition through five distinct subcategories: observing, describing, acting with awareness, non-judging of inner experience, and non-reactivity to inner experience. We compare the pre- and post-intervention scores to assess changes in trait mindfulness.

Finally, the participants completed the second version of the Multidimensional Assessment of Interoceptive Awareness (MAIA-2)~\cite{mehling2018multidimensional}. This assessment is utilized to evaluate the subjective perception of the individual regarding internal bodily sensations. It captures the multidimensional nature of interoception through domains such as the noticing of somatic sensations, the regulation of psychological distress through somatic attention, and the emotional response to bodily states. The administration of MAIA-2 facilitates a deeper understanding of the interaction between the somatic awareness of the individual and the specific cognitive demands of the different meditative paradigms, thereby providing a robust psychological framework for the observed EEG data.

\subsubsection{Psychological Assessment Outcomes}
This section presents the quantitative outcomes of the psychological assessments administered during the pre-intervention and post-intervention sessions. Specifically, we evaluate the longitudinal shifts in participant responses across three validated instruments: the Perceived Stress Scale (PSS) to measure stress reduction, the Multidimensional Assessment of Interoceptive Awareness version 2 (MAIA-2) to assess interoceptive body awareness, and the Five Facet Mindfulness Questionnaire: Short-Form (FFMQ-SF) to quantify trait mindfulness. The subsequent paragraphs detail the statistical variations and psychometric profile shifts for the SA, HK, and BF groups to establish the psychological efficacy of the respective interventions.

\begin{table}[ht]
  \centering
  \small
  \caption{\small Questionnaire descriptive statistics (mean $\pm$ standard deviation) by arm for the strict cross-instrument paired cohort. The scores for the MAIA-2 and FFMQ-sf represent the average of the total summed scores for each group. Note: Questionnaire data were available for 73 of 74 participants at baseline. Longitudinal analyses were restricted to 43 participants with complete pre- and post-intervention data across all three questionnaires.}
  \label{tab:questionnaires-item-mean-intersection}
  \begin{tabular}{@{}llccccc@{}}
    \toprule
    & & \multicolumn{2}{c}{\textbf{Unpaired ($n=73$)}} & \multicolumn{3}{c}{\textbf{Paired Cohort ($n=43$)}} \\
    \cmidrule(lr){3-4} \cmidrule(l){5-7}
    \textbf{Measure} & \textbf{Group} & $n$ & \textbf{Pre} & $n$ & \textbf{Pre} & \textbf{Post} \\
    \midrule
    \multirow{3}{*}{\textbf{PSS}~$^\downarrow$}
    & SA & 26 & $16.2 \pm 4.6$ & 16 & $17.7 \pm 5.6$ & $13.8 \pm 6.0$ \\
    & HK & 31 & $18.4 \pm 5.8$ & 19 & $19.0 \pm 6.0$ & $15.6 \pm 5.0$ \\
    & BF & 16 & $19.2 \pm 6.3$ &  8 & $20.5 \pm 7.4$ & $15.1 \pm 7.1$ \\
    \midrule
    \multirow{3}{*}{\textbf{MAIA-2}~$^\uparrow$}
    & SA & 26 & $2.84 \pm 0.66$ & 16 & $2.88 \pm 0.65$ & $3.44 \pm 0.25$ \\
    & HK & 31 & $2.58 \pm 0.56$ & 19 & $2.65 \pm 0.44$ & $3.10 \pm 0.57$ \\
    & BF & 16 & $2.75 \pm 0.66$ &  8 & $2.86 \pm 0.79$ & $3.65 \pm 0.53$ \\
    \midrule
    \multirow{3}{*}{\textbf{FFMQ-SF}~$^\uparrow$}
    & SA & 26 & $3.22 \pm 0.46$ & 16 & $3.24 \pm 0.53$ & $3.36 \pm 0.47$ \\
    & HK & 31 & $3.05 \pm 0.45$ & 19 & $3.20 \pm 0.39$ & $3.38 \pm 0.33$ \\
    & BF & 16 & $3.27 \pm 0.51$ &  8 & $3.22 \pm 0.52$ & $3.55 \pm 0.49$ \\
    \bottomrule
  \end{tabular}
\end{table}

\begin{figure}[ht]
  \centering
  \includegraphics[width=\linewidth]{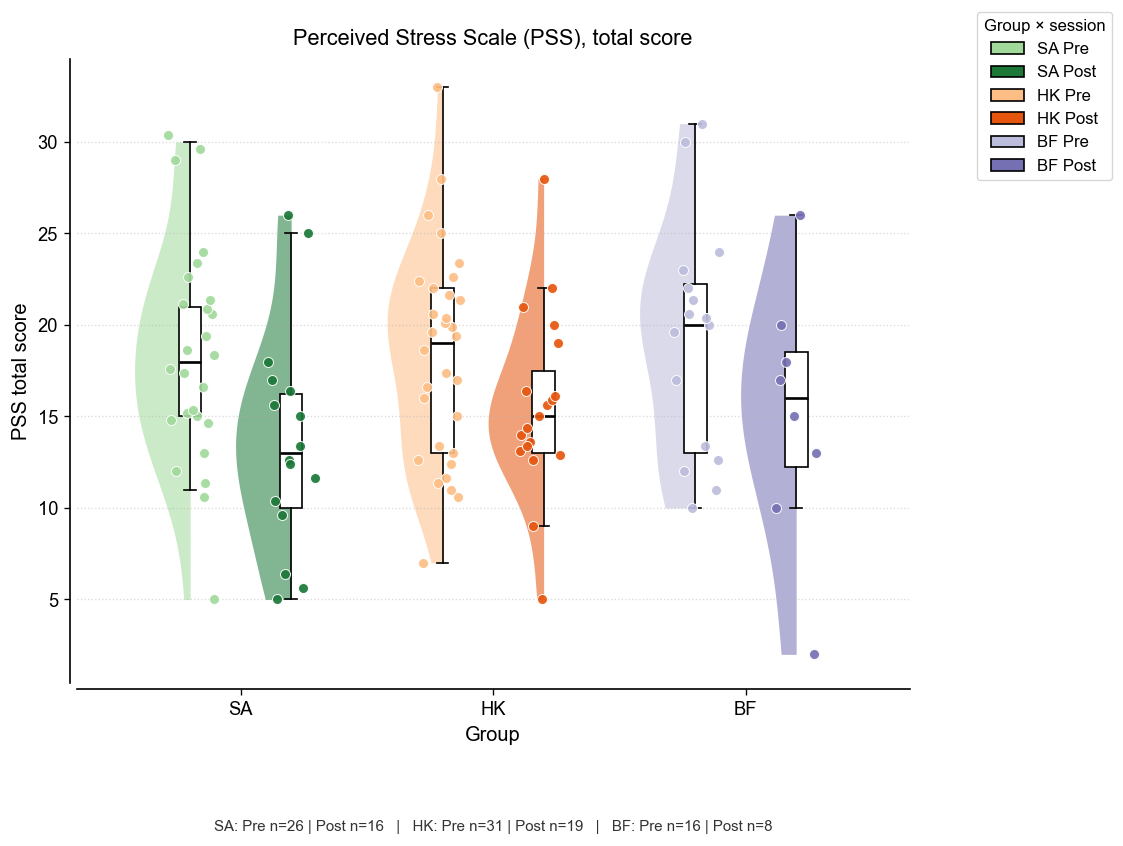}
  \caption{Total PSS scores at baseline and follow-up for SA, HK, and BF.}
  \label{fig:pss-raincloud}
\end{figure}

\paragraph{Quantitative Changes in the Perceived Stress Scale}
Figure~\ref{fig:pss-raincloud} illustrates the total scores of the Perceived Stress Scale (PSS) for the SA, HK, and BF groups during both the pre-intervention and post-intervention sessions. Within the plots, each data point represents the raw PSS total score of an individual participant, accompanied by a probability density curve and a box plot to indicate the data distribution.

Table~\ref{tab:questionnaires-item-mean-intersection} details the differences in scores between the two sessions. The comparative analysis between the pre-intervention and post-intervention stages relies on a paired cohort of 43 participants who completed the questionnaire assessments at both time points. Although the table includes an unpaired column detailing the overall pre-intervention baseline results for the entire initial participant pool, all longitudinal comparisons and subsequent quantitative findings discussed throughout this section, including the analyses of the MAIA-2 and FFMQ-SF questionnaires, derive exclusively from this paired cohort. Within this cohort, the mean PSS total scores demonstrate a consistent reduction from the pre-intervention session to the post-intervention session across all groups. Specifically, the mean total score for the SA group significantly decreased ($p < 0.05$) from 17.7 at pre-intervention to 13.8 at post-intervention. Furthermore, the mean total score for the HK group significantly decreased from 19.0 to 15.6, and the corresponding score for the BF group non-significantly decreased from 20.5 to 15.1. 

\begin{figure}[ht]
  \centering
  \includegraphics[width=\linewidth]{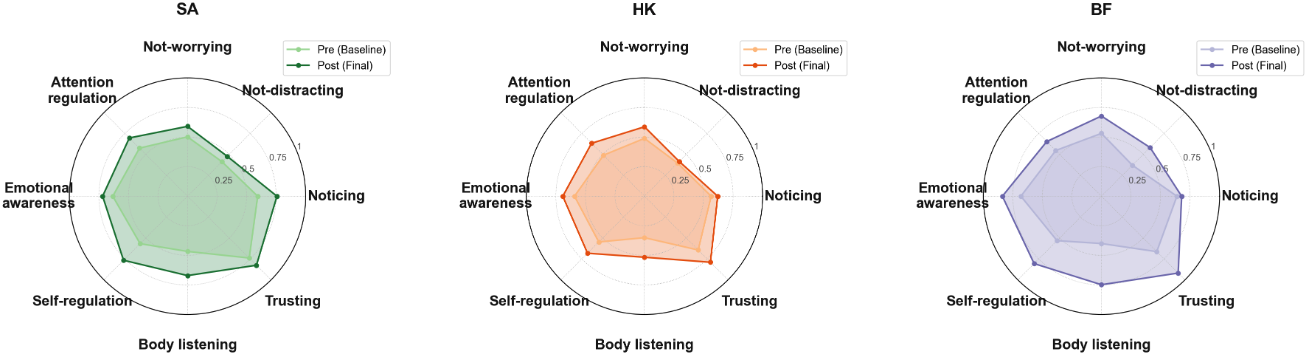}
  \caption{Mean MAIA-2 scores at baseline and follow-up for SA, HK, and BF on eight domains.}
  \label{fig:maia-radar-mean}
\end{figure}

\paragraph{Quantitative Changes in the Multidimensional Assessment of Interoceptive Awareness}
The Multidimensional Assessment of Interoceptive Awareness version 2 (MAIA-2) evaluates bodily awareness across eight subscales. These include Noticing (the awareness of comfortable, uncomfortable, and neutral sensations), Not-Distracting (the tendency not to ignore pain or discomfort), Not-Worrying (the absence of emotional distress when experiencing discomfort), Attention Regulation (the ability to sustain and control attention on bodily sensations), Emotional Awareness (the recognition of connections between bodily sensations and emotional states), Self-Regulation (the ability to manage psychological distress by attending to bodily sensations), Body Listening (the active receptivity to physiological signals for insight), and Trusting (the perception of the body as safe and trustworthy). Figure~\ref{fig:maia-radar-mean} presents normalized radar plots for the SA, HK, and BF groups, illustrating shifts in these subscale profiles from pre- to post-intervention.

Subscale profiles varied across the study groups after the intervention. The SA group improved across most domains, particularly in Not-Distracting, Trusting, and Attention Regulation. The HK group showed similar overall gains, primarily in Attention Regulation, Self-Regulation, and Emotional Awareness. The BF group improved in Self-Regulation, Emotional Awareness, and Trusting, but showed negligible change in Noticing. Table~\ref{tab:questionnaires-item-mean-intersection} details these quantitative changes. Overall MAIA-2 item means non-significantly increased for all groups: from $2.88$ to $3.44$ for the SA group, from $2.65$ to $3.10$ for the HK group, and from $2.86$ to $3.65$ for the BF group.

\begin{figure}[ht]
  \centering
  \includegraphics[width=\linewidth]{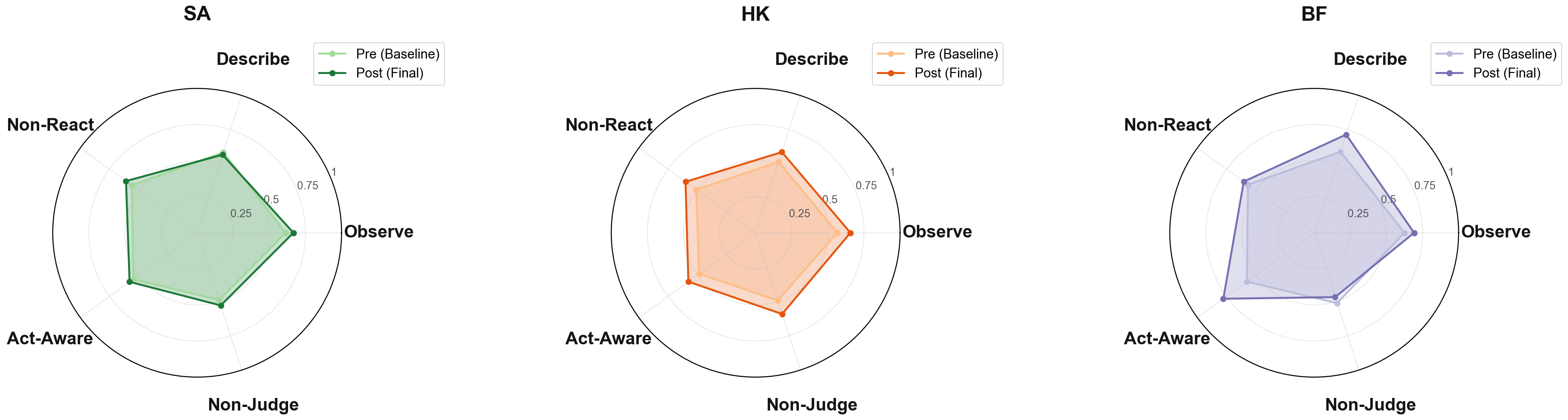}
  \caption{Mean FFMQ-SF scores at baseline and follow-up for SA, HK, and BF on five facets.}
  \label{fig:ffmq-radar-mean}
\end{figure}

\paragraph{Quantitative Changes of the Five Facet Mindfulness Questionnaire: Short-Form}
The Five Facet Mindfulness Questionnaire: Short-Form (FFMQ-SF) assesses trait mindfulness across five distinct subscales using a five-point Likert scale, ranging from 1 (never or very rarely true) to 5 (very often or always true). These subscales encompass Observing (the conscious perception of internal and external sensory experiences), Describing (the ability to clearly label internal experiences with words), Acting with Awareness (the maintenance of focused attention on current activities without operating on autopilot), Non-judging of Inner Experience (the adoption of a non-evaluative attitude toward personal thoughts and emotions), and Non-reactivity to Inner Experience (the capacity to allow thoughts and emotions to pass without getting entangled or reacting impulsively). Similar to the visualization approach employed for the MAIA-2, Figure~\ref{fig:ffmq-radar-mean} presents normalized radar plots for the SA, HK, and BF groups, depicting relative shifts in these facet profiles from pre-intervention to post-intervention.

Distinct variations in facet profiles are observable across the study groups following the intervention. In the SA group, post-intervention scores demonstrate minor improvements across all facets expect slightly decrease in Describe facet. The HK group exhibits a general expansion of the mindfulness profile, with prominent enhancements in all five facets. Conversely, the BF group displays a distinct pattern characterized by notable increases in Describing, Observing, Acting with Awareness, and Non-reactivity, coupled with a discernible decrease in the Non-judging facet. These visual profile shifts are quantitatively supported by the data in Table~\ref{tab:questionnaires-item-mean-intersection}, which presents the average of the total summed scores for each group. Specifically, the mean score for the SA group increases from 3.24 at pre-intervention to 3.36 at post-intervention. Furthermore, the mean score for the HK group non-significantly increases from 3.20 to 3.38, and the mean score for the BF group rises from 3.22 to 3.55.

%\begin{table}[ht]
  %\centering
  %\small
  %\caption{Temporary for comparison, this is Hamzeh's result table 9}
  %\label{tab:questionnaires-item-mean-intersection}
  %\begin{tabular}{@{}llccccc@{}}
    %\toprule
    %& & \multicolumn{2}{c}{\textbf{Unpaired (All)}} & \multicolumn{3}{c}{\textbf{Paired Cohort ($n=38$)}} \\
    %\cmidrule(lr){3-4} \cmidrule(l){5-7}
    %\textbf{Measure} & \textbf{Group} & $n$ & \textbf{Pre} & $n$ & \textbf{Pre} & \textbf{Post} \\
    %\midrule
    %\multirow{3}{*}{\textbf{PSS}~$^\downarrow$}
    %& SA & 26 & $16.2 \pm 4.6$ & 15 & $16.8 \pm 4.8$ & $19.0 \pm 6.7$ \\
    %& HK & 31 & $16.7 \pm 6.0$ & 16 & $17.0 \pm 6.9$ & $17.1 \pm 5.2$ \\
    %& BF & 16 & $17.4 \pm 5.9$ &  7 & $20.0 \pm 5.8$ & $14.9 \pm 5.1$ \\
    %\midrule
    %\multirow{3}{*}{\textbf{MAIA-2}~$^\uparrow$}
    %& SA & 26 & $2.87 \pm 0.88$ & 15 & $2.96 \pm 1.02$ & $3.07 \pm 0.55$ \\
    %& HK & 31 & $3.01 \pm 0.47$ & 16 & $3.08 \pm 0.54$ & $2.91 \pm 0.50$ \\
    %& BF & 16 & $3.01 \pm 0.53$ &  7 & $3.19 \pm 0.52$ & $2.78 \pm 0.51$ \\
    %\midrule
    %\multirow{3}{*}{\textbf{FFMQ-SF}~$^\uparrow$}
    %& SA & 26 & $15.29 \pm 2.59$ & 15 & $15.73 \pm 2.09$ & $15.20 \pm 1.66$ \\
    %& HK & 31 & $15.66 \pm 2.41$ & 16 & $14.78 \pm 1.97$ & $15.96 \pm 2.72$ \\
    %& BF & 16 & $15.84 \pm 2.26$ &  7 & $16.00 \pm 2.04$ & $16.60 \pm 2.68$ \\
    %\bottomrule
  %\end{tabular}
%\end{table}

\section{Detailed Methodology, Data Quality and Data Processing}

\subsection{Quantitative EEG Quality Assessment}
\label{app:eeg-qc}

\paragraph{Data and reporting.}
We evaluated channel-level and spectral EEG quality using measures informed by the PREP and HAPPE frameworks~\citep{bigdelyshamlo2015prep,gabarddurnam2018happe}. Raw-recording QC included 118 sessions (74 pre-intervention and 44 post-intervention). Benchmark-input QC covered 472 minimally preprocessed task segments, comprising four tasks per session: \texttt{restCE01}, \texttt{Medita}, \texttt{restCE02}, and \texttt{slMedita}. The QC results characterize the raw recordings and minimally preprocessed benchmark inputs, rather than the separately released fully cleaned EEG data. Table~\ref{tab:eeg-qc} summarizes the results as medians and first and third quartiles (Q1 and Q3) across sessions or task segments.

\paragraph{Raw-recording measures.} Channel QC was performed on a 1-Hz high-pass-filtered copy of each 64-channel recording before interpolation, without additional re-referencing or ICA. We used \texttt{clean\_rawdata} with \texttt{FlatlineCriterion}=5\,s, \texttt{ChannelCriterion}=0.8, and \texttt{LineNoiseCriterion}=4; internal high-pass filtering, ASR, and window rejection were disabled~\citep{DelormeMakeig2004}. The good-channel percentage denotes channels meeting these criteria. For local channel similarity, we computed Pearson correlations   over the pass-filtered recording between each channel and its four nearest neighbours in the standard 10-05 system, then took the median across neighbours and channels.

Spectral measures used the unfiltered recording, with Welch spectra estimated using 4-s Hamming windows and 50\% overlap. The 60-Hz ratio compares mean PSD at 59.5--60.5\,Hz with that over the combined 55--58 and 62--65\,Hz bands~\citep{tadel2019meg}. These measures describe spectral characteristics rather than directly quantifying non-neural signal.

\paragraph{Benchmark-input measures.} Each task segment contained 64 channels sampled at 250\,Hz and was assessed before runtime normalization. Within non-overlapping 4-s windows, a channel was flagged if its peak-to-peak amplitude exceeded 150 or its population SD fell below 0.5, in stored amplitude units. Either condition triggered a flag, with each channel-window pair counted only once. For each task segment, we averaged the percentage of flagged channels across windows, excluding any incomplete final window. These segment-level percentages were then summarized across the 472 task segments. Flags were descriptive and did not remove data. We additionally calculated each channel's population SD over the entire task segment and took the median across channels to obtain the segment-level amplitude variability.

\begin{table}[H]
\centering
\small
\setlength{\tabcolsep}{5pt}
\caption{Quantitative EEG quality assessment. Values are median [Q1, Q3] across 118 raw sessions or 472 task segments. Metric definitions and exceptions are given above.}
\label{tab:eeg-qc}
\begin{tabular}{@{}llrl@{}}
\toprule
Data & Metric & $N$ & Median [Q1, Q3] \\
\midrule
Raw & Good channels (\%) & 118 & 98.4 [96.9, 100.0] \\
Raw & Four-neighbour correlation & 118 & 0.884 [0.845, 0.907] \\
Raw & 60-Hz ratio (dB) & 118 & 0.08 [$-0.18$, 0.37] \\
\midrule
Benchmark & Flagged channel-window pairs (\%) & 472 & 0.56 [0.10, 2.94] \\
Benchmark & Channel SD (stored units) & 472 & 10.75 [8.89, 12.89] \\
\bottomrule
\end{tabular}
\end{table}

\subsection{Preprocessing Details}
\label{data_preprocess}

 For data in cleaned EEG in Section~\ref{data_struct}, the High-pass conditioning utilized a 1 Hz zero-phase Butterworth filter. Zapline-plus removed 60 Hz interference via spectral integration without distorting neural oscillations~\citep{klug2022zaplineplus}. ASR was configured with a 25-standard-deviation burst detection threshold and 0.2 maximum bad-channel tolerance. Channels flagged by ASR were interpolated using spherical splines to maintain the 64-channel manifold~\citep{chang2019evaluation}. A zero-valued reference was appended before common average re-referencing across all electrodes plus the FCz channel, totaling 65 channels. ICLabel components were rejected if the artifact probability reached $\ge 0.9$. For the training data used for Benchmark tasks, the algorithm simply preprocessed the data with a spatial correlation threshold of 0.9 and a line-noise criterion of 4 standard deviations.

\begin{wrapfigure}[17]{r}{0.48\textwidth}
 \centering
  \includegraphics[width=0.45\textwidth]{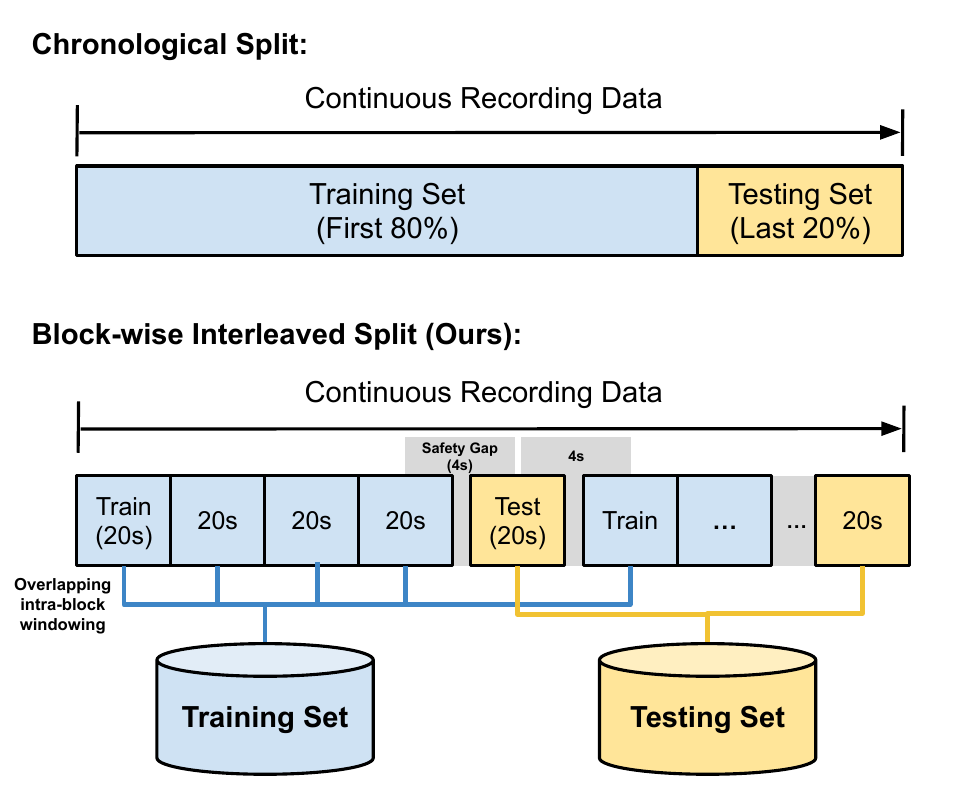}  
  \caption{\small Chronological split and block-wise interleaved split(Ours) for intra-subject data splitting. }
  \label{fig:intra_split} 
\end{wrapfigure}

\subsection{Model Training Details}
\label{training_detail}

\paragraph{Temporal Partitioning And Class Balancing} At $250 \text{ Hz}$, the recordings yield $64 \times 1000$ feature matrices across 64 channels. For intra-subject evaluations, the data are partitioned into 20-second blocks, alternating between training and testing sets in an $80\%$/$20\%$ ratio. Unlike standard chronological splitting, this block-wise strategy is utilized exclusively in the intra-subject setting to account for non-stationary temporal dynamics throughout the session. To prevent data leakage between adjacent blocks, 4-second safety gaps are inserted (Figure~\ref{fig:intra_split}). Additionally, to address duration discrepancies between resting and meditation states, a dynamic weighted random sampler is implemented to ensure a 1:1 class exposure per mini-batch during training.

\paragraph{Training traditional SVMs on FBCSP and PSD extracted features.} For FBCSP, we use a sampling frequency of 250 Hz, number of components $k = 4$ which is the number of features extracted from every frequency band, we filter the EEG reading for 9 bands, namely, (4,8), (8,12), (12,16), (16,20), (20,24), (24,28), (28,32), (32,36), and (36,40) Hz each. The input dimension of the EEG reading to algorithm is $N \times C \times T$, and the output dimension is $N\times Bk$, where $N$ is the number of trials which varies based on the evaluation, $C=64$ is the number of channels, and $T=1000$ is the time steps (number of snapshots taken during EEG reading), $B=9$ represents the 9 bands extracted from each EEG reading, and $k=4$ is the number of features extracted from each band using the band's specific filter $w_b$. For the PSD feature extraction, we use the same sampling frequency, but different frequency bands of delta (1, 4), theta (4, 8), alpha (8, 13), beta (13, 30), and gamma (30, 45) and the output dimension of the features is $N \times BC$ because PSD measures frequency power for each channel and uses those as features. We train an SVM on top of features extracted from both of these methods, the SVM has a linear kernel and parameter $C = 1.0$.

% \vspaces
\subsection{Benchmark Task Detailed Methodology}
\label{benchmark_detail}

\paragraph{Data and Cohort Specifications for Task 1.} This paragraph supplements the methodological discussion in subsection Task 1: Cognitive State Decoding. The initial session includes the complete cohort ($N=74$), avoiding the attrition observed in the post-intervention phase ($N=44$). Continuous recordings comprise a four-minute resting baseline and an eight-minute meditation task. The global model is trained on all 74 subjects to establish a comparison baseline for the subgroup-specific models (Table \ref{tab:task1_groups_full}). The resting baseline was always recorded before silent meditation. Note that Task 1 decoding performance should therefore be interpreted in the context of this fixed recording order, as time-related changes may also contribute to the observed differences.

\paragraph{Evaluation Protocol for Task 2}
This paragraph supplements the methodological discussion subsection Task 2 uses stratified 5-fold inter-subject cross-validation to evaluate meditation technique classification. Since each participant practices only one technique, intra-subject classification is not applicable. Although leave-one-subject-out (LOSO) evaluation is possible, each test fold contains only one class, making fold-level multi-class metrics unsuitable. We therefore use subject-disjoint stratified folds to evaluate generalization across unseen participants.

\begin{figure}[ht]
    \centering
    \includegraphics[width=\linewidth]{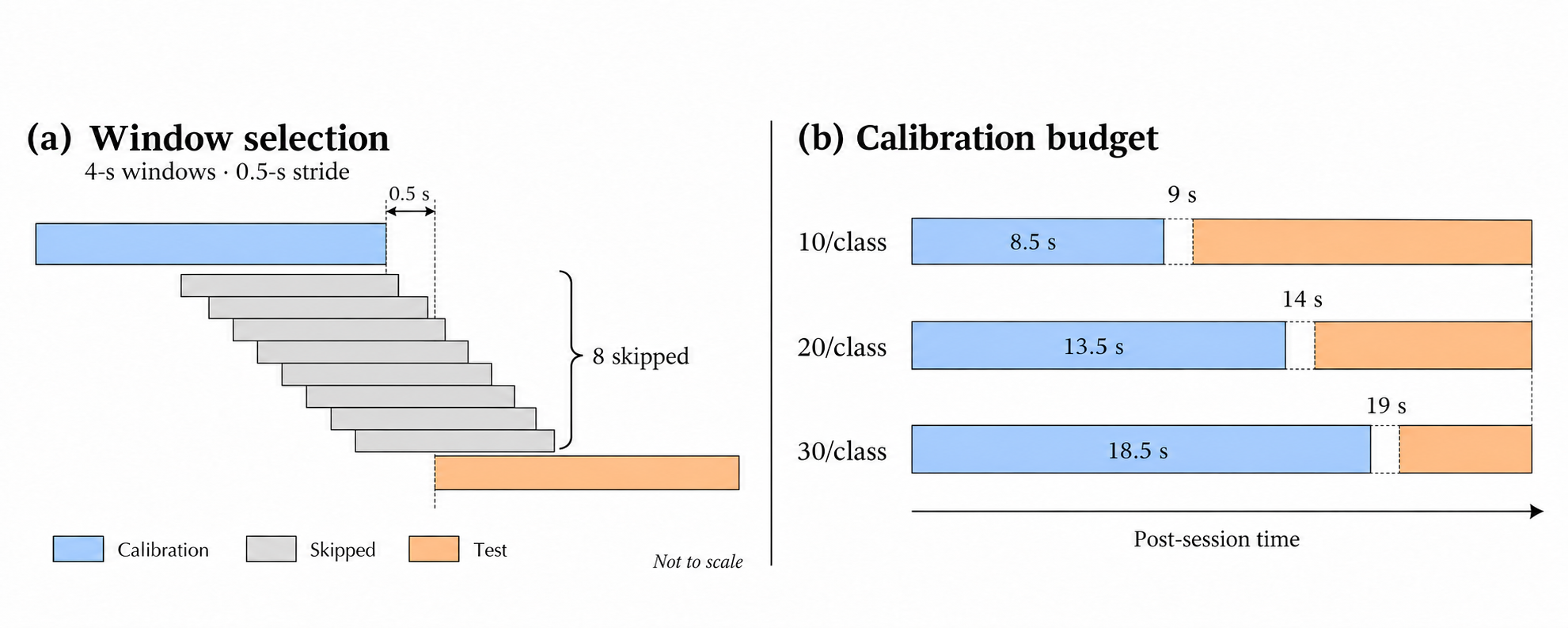}
    \caption{\small \textbf{Task 3 calibration and evaluation split.} \textbf{(a)} Window selection with a 0.5-s gap between calibration and evaluation data. \textbf{(b)} Calibration coverage and test onset for different per-class shot budgets. Blue, grey, and orange indicate calibration, excluded windows, and evaluation, respectively. }
    \label{fig:task3_calibration_split}
\end{figure}

\paragraph{Calibration Protocols for Task 3.} This paragraph supplements the methodology described in subsection~\ref{task3_label}. Task 3 evaluated within-subject transfer across the six-week intervention using 44 participants with paired pre- and post-intervention sessions. Models were initialized from subject-specific Task 1 checkpoints trained on pre-intervention data using a chronological split. Post-intervention recordings from \texttt{restCE01} and \texttt{slMedita} were segmented into 4-s windows with a 0.5-s stride, representing the two classification classes. For $N$-shot calibration, the first $N$ windows from each class were selected, yielding $2N$ calibration windows in total. As shown in Figure~\ref{fig:task3_calibration_split}(a), the next eight candidate windows per class were excluded before selecting the remaining windows for evaluation. This produced a 0.5-s gap between the last calibration window and the first evaluation window, ensuring that the two subsets shared no EEG samples.

Accounting for overlapping windows, the 10-, 20-, and 30-shot calibration sets covered 8.5, 13.5, and 18.5 s per class, respectively, with testing beginning at 9, 14, and 19~s relative to the start of each class recording (Figure~\ref{fig:task3_calibration_split}(b)). The remaining windows were used for testing, with the test onset determined by the calibration budget. Zero-shot evaluation used all eligible Post windows. For each participant and calibration budget, the data split was fixed across models and fine-tuning strategies. No separate post-intervention validation set was used, model adaptation relied solely on the calibration windows, with performance evaluated on the held-out test windows. Each window was independently standardized per channel using its temporal mean and standard deviation. Test-set sizes differed only slightly across calibration budgets. Each additional 10 shots per class shifted the test onset by 5 s within each recording, reducing the total test set by only 20 windows.

\vspace{-0.1in}

\subsection{Implementation Details}

\paragraph{Hardware, Software, and Parallelization} All experiments were conducted on a server equipped with eight NVIDIA RTX A5000 GPUs (24 GB VRAM each). Models were implemented in PyTorch 2.6.0 with CUDA 11.8. Each training run was launched as an independent single-GPU process and assigned to a dedicated device via a custom scheduling script, enabling up to eight concurrent runs without inter-GPU communication overhead. All computations were performed in standard float32 precision.

\paragraph{Optimization and Hyperparameter Search} All models were optimized using Adam ($\beta_1 = 0.9, \beta_2 = 0.999$) with cross-entropy loss and a constant learning rate throughout training. Hyperparameters were selected via Optuna with 50 trials per model--task combination, maximizing validation balanced accuracy; unpromising trials were pruned using the Median Pruner (warm-up steps = 10). The search covered learning rate $\in [10^{-5}, 10^{-3}]$ (log-uniform; restricted to $[10^{-5}, 5 \times 10^{-4}]$ for Conformer), weight decay $\in [10^{-5}, 10^{-2}]$, dropout rate $\in [0.05, 0.7]$, temporal kernel size, and model-specific architectural parameters (e.g., embedding size and attention depth for Conformer). For Tasks 1 and 2, final models were trained for up to 250 epochs with a batch size of 32 and early stopping (patience = 50 epochs) based on validation balanced accuracy. For Task 3, we evaluated two adaptation strategies: full fine-tuning, which updates all learnable parameters, and BN-and-head fine-tuning, which updates only the BatchNorm affine parameters and the final classification head. Both strategies used 30 adaptation epochs and a batch size of four.

To ensure reproducibility, all random seeds across PyTorch, NumPy, and Optuna samplers were fixed to a designated value, and deterministic algorithms were enforced where supported.

% ========================================================== %
\section{Additional Experimental Results}

\subsection{Benchmark Task1 Supplementary Material}
\label{app:task1} 
This supplementary section provides a multidimensional analysis of the decoding performance for Benchmark Task 1, extending beyond standard aggregate metrics to dissect the critical bottlenecks in EEG-based classification. Specifically, we present detailed investigations into four key aspects: (1) the profound performance discrepancy between subject-specific and cross-subject evaluation protocols, (2) cohort-specific biases and cross-group generalization dynamics across three distinct sub-populations (SA, HK, and BF), (3) the impact of temporal non-stationarity evaluated through a rigorous chronological-split protocol versus standard block-randomized cross-validation, and (4) subject-level heterogeneity, which isolates specific algorithmic vulnerabilities to traditionally challenging participants.

The primary objective of including these granular analyses is to expose the hidden variances, temporal drifts, and dataset biases that are frequently obscured by conventional mean-accuracy reporting. For researchers utilizing this dataset, these results serve as a diagnostic map of the data's inherent complexities. They demonstrate that idealized intra-subject evaluations represent a theoretical upper bound, whereas inter-subject and temporal-causal protocols reflect true real-world robustness.

Future work could evaluate EEG foundation models, such as LaBraM and REVE~\citep{jiang2024large,el2026reve}, on the L-FAME benchmark. Comparing these pretrained models with our task-specific baselines under the same evaluation protocols would help determine whether large-scale EEG pretraining improves cross-subject generalization and decoding performance for challenging participants.

\paragraph{Performance Degradation for Cross-Subject and Superior Generalization of Convolutional Architectures}

The experimental results from Table~\ref{tab:task1_detailed_split} reveal a substantial performance discrepancy between subject-specific and cross-subject evaluation strategies. In the Block-wise and Chrono-wise intra-subjects splitting settings, most models achieve high decoding performance, with EEGNet reaching 99.2\% and 96.8\% AUC, respectively. However, when transitioning to Inter-Subject and LOSO protocols, all models experience a drastic decline in performance. This significant drop underscores the severe challenge posed by inter-subject variability and distribution shifts in EEG signals, indicating that models heavily rely on subject-specific temporal features when trained and evaluated on individualized data splits.

 Comparing traditional machine learning approaches with deep neural networks reveals a stark contrast in cross-domain generalization capabilities. The FBCSP + SVM pipeline achieves near-perfect metrics in the block-wise setting, recording an AUC of 99.9\% and an accuracy of 99.4\%. Nevertheless, its performance drops to near-chance levels, specifically 55.8\% AUC and 53.8\% accuracy, under the Inter-Subject protocol. Same thing happens when the FBCSP and PSD feature extraction methods are combined with an SVM in the LOSO setting. The AUC for FBCSP + SVM drops to 58.2\% while the PSD + SVM drop to 61.3\%. This shows that these methods struggle when presented with data that wasn't included when fitting the SVM, and hence they have low ability to generalize, making them best suited to use for Intra-Subject evaluations.
 
In contrast to traditional methods, deep learning models such as ShallowConvNet and EEG-Conformer demonstrate superior generalizability and maintain a much stronger baseline in cross-subject scenarios (inter-subjects and LOSO). It should be noted that EEG-Conformer explicitly incorporates a shallow convolutional module as its initial feature extraction head before applying its global self-attention mechanism. The strong performance of both architectures indicates that CNN learned spatial-temporal representations offer greater robustness to subject-specific noise than traditional hand-crafted spatial filters. Furthermore, the observation that a relatively simple architecture like ShallowConvNet achieves highly competitive results suggests that deeper convolutional models may be highly susceptible to overfitting when deployed across distinct individuals. Consequently, utilizing fewer convolutional layers appears to act as an implicit regularization mechanism, successfully preventing the network from memorizing idiosyncratic, subject-specific artifacts and thereby preserving cross-population generalizability.

\begin{table}[ht]
  \centering
  \small 
  \renewcommand{\arraystretch}{1.2}
  \setlength{\tabcolsep}{3.5pt} 
  \caption{\small \textbf{Task 1: Decoding Performance Across Protocols.} Detailed metrics for each evaluation strategy. Results are reported as Mean $\pm$ SD across subjects ($N=74$). \textbf{Bold} indicates the best performance within each strategy group.}
  \label{tab:task1_detailed_split}
  \begin{tabular}{@{} l l cccc @{}}
    \toprule
    \textbf{Strategy} & \textbf{Model} & \textbf{AUC (\%)} & \textbf{Acc (\%)} & \textbf{BAcc (\%)} & \textbf{F1 (\%)} \\
    \midrule
    \multirow{6}{*}{\begin{tabular}[c]{@{}l@{}}\textbf{Intra-Subject}\\ (block-wise)\end{tabular}}
    & PSD + SVM & 97.1{\scriptsize $\pm$4.0}  & 92.9{\scriptsize $\pm$7.2}  & 92.1{\scriptsize $\pm$8.0} & 92.0{\scriptsize $\pm$8.1}
    \\
    & FBCSP + SVM    & \textbf{99.9{\scriptsize $\pm$0.2}} & \textbf{99.4{\scriptsize $\pm$1.5}} & \textbf{99.3{\scriptsize $\pm$1.9}} & \textbf{99.4{\scriptsize $\pm$1.7}}
    
    \\

    & ShallowConvNet & 97.2{\scriptsize $\pm$4.4} & 92.6{\scriptsize $\pm$6.1} & 90.6{\scriptsize $\pm$7.7} & 91.3{\scriptsize $\pm$7.2} \\
    & DeepConvNet      & 97.0{\scriptsize $\pm$8.5} & 88.0{\scriptsize $\pm$15.2} & 86.7{\scriptsize $\pm$14.2} & 85.5{\scriptsize $\pm$18.1} \\
    & EEGNet           & 
    99.2{\scriptsize $\pm$2.1} & 96.2{\scriptsize $\pm$4.6} & 95.4{\scriptsize $\pm$6.0} & 95.6{\scriptsize $\pm$5.5} \\
    & EEG-Conformer    & 97.0{\scriptsize $\pm$4.9} & 92.3{\scriptsize $\pm$6.9} & 90.3{\scriptsize $\pm$9.1} & 90.8{\scriptsize $\pm$8.6} \\
    \midrule
    
     \multirow{6}{*}{\begin{tabular}[c]{@{}l@{}}\textbf{Intra-Subject}\\ (chrono-wise)\end{tabular}} 
     & PSD + SVM & 94.3{\scriptsize $\pm$7.0}  & 89.0{\scriptsize $\pm$8.6}  & 87.0{\scriptsize $\pm$10.6} & 87.1{\scriptsize $\pm$10.4}
    \\
    & FBCSP + SVM    & 
    \textbf{99.5{\scriptsize $\pm$2.1}} & \textbf{98.5{\scriptsize $\pm$5.0}} & \textbf{97.9{\scriptsize $\pm$7.6}} & \textbf{97.8{\scriptsize $\pm$8.0}} \\

    & ShallowConvNet & 92.8{\scriptsize $\pm$8.7} & 86.9{\scriptsize $\pm$9.0} & 82.4{\scriptsize $\pm$12.6} & 83.2{\scriptsize $\pm$12.5} \\
    & DeepConvNet      & 95.0{\scriptsize $\pm$11.2} & 83.7{\scriptsize $\pm$14.5} & 80.0{\scriptsize $\pm$16.4} & 78.4{\scriptsize $\pm$19.5} \\
    & EEGNet           & 
    96.8{\scriptsize $\pm$5.5} & 91.2{\scriptsize $\pm$8.9} & 88.9{\scriptsize $\pm$11.4} & 89.2{\scriptsize $\pm$11.1} \\
    & EEG-Conformer    & 92.5{\scriptsize $\pm$9.3} & 86.6{\scriptsize $\pm$9.3} & 82.3{\scriptsize $\pm$12.3} & 83.2{\scriptsize $\pm$12.2} \\
    \midrule
    
    \multirow{6}{*}{\textbf{Inter-Subject}} 
     & PSD + SVM & 59.4{\scriptsize $\pm$3.8}   & 57.7{\scriptsize $\pm$4.2}  & 56.9{\scriptsize $\pm$2.9} & 55.6{\scriptsize $\pm$3.4}
    \\
    & FBCSP + SVM    & 55.8{\scriptsize $\pm$6.2} & 53.8{\scriptsize $\pm$4.7} & 53.7{\scriptsize $\pm$3.6} & 52.1{\scriptsize $\pm$3.9} 
    \\
    & ShallowConvNet & 66.5{\scriptsize $\pm$3.9} & \textbf{64.9{\scriptsize $\pm$4.9}} & \textbf{61.3{\scriptsize $\pm$3.1}} & \textbf{60.7{\scriptsize $\pm$3.7}} \\
    & DeepConvNet      & 66.2{\scriptsize $\pm$5.3} & 62.8{\scriptsize $\pm$5.3} & 59.5{\scriptsize $\pm$2.8} & 59.0{\scriptsize $\pm$3.6} \\
    & EEGNet           & 66.5{\scriptsize $\pm$3.5} & 60.6{\scriptsize $\pm$5.3} & 60.9{\scriptsize $\pm$3.7} & 59.0{\scriptsize $\pm$4.5} \\
    & EEG-Conformer    & \textbf{66.9{\scriptsize $\pm$4.7}} & 64.3{\scriptsize $\pm$5.0} & 61.1{\scriptsize $\pm$3.1} & 60.6{\scriptsize $\pm$3.7} \\
    \midrule
    \multirow{6}{*}{\textbf{LOSO}} 
     & PSD + SVM & 61.3{\scriptsize $\pm$4.4}  & 56.7{\scriptsize $\pm$2.3}  & 57.9{\scriptsize $\pm$2.8} & 55.6{\scriptsize $\pm$12.5}
    \\
    & FBCSP + SVM    & 58.2{\scriptsize $\pm$21.4} & 61.7{\scriptsize $\pm$11.8} & 52.7{\scriptsize $\pm$10.8}  & 47.6{\scriptsize $\pm$12.0}  \\
   
    & ShallowConvNet & \textbf{70.4{\scriptsize $\pm$20.8}} & \textbf{63.9{\scriptsize $\pm$16.7}} & \textbf{61.5{\scriptsize $\pm$15.0}} & \textbf{56.7{\scriptsize $\pm$17.9}} \\
    & DeepConvNet      & 68.4{\scriptsize $\pm$26.8} & 61.3{\scriptsize $\pm$20.4} & 59.8{\scriptsize $\pm$18.2} & 54.6{\scriptsize $\pm$20.7} \\
    & EEGNet           & 66.6{\scriptsize $\pm$27.2} & 58.4{\scriptsize $\pm$21.2} & 59.8{\scriptsize $\pm$17.9} & 52.5{\scriptsize $\pm$21.9} \\
    & EEG-Conformer    & 67.5{\scriptsize $\pm$24.3} & 62.7{\scriptsize $\pm$17.8} & 60.0{\scriptsize $\pm$15.8} & 55.3{\scriptsize $\pm$18.3} \\
    \bottomrule
  \end{tabular}
\end{table}

\paragraph{Evaluating Cohort Biases And Cross-Group Generalization Dynamics} To investigate how different population distributions affect model robustness, the dataset is partitioned into three distinct sub-datasets: SA with $n=27$, HK with $n=31$, and BF with $n=16$. The fundamental purpose of this group-specific experiment is to isolate cohort characteristics and reveal how internal data variance and sample size fundamentally influence model generalization. For the intra-subject protocol, the displayed metrics represent the average decoding accuracy computed across all individuals strictly within their respective group. Conversely, for the inter-subject and leave-one-subject-out protocols, the dataset is completely re-partitioned, meaning models are independently trained and tested exclusively within each of the three isolated subsets. While the intra-subject results display uniformly excellent performance across all groups, with architectures like EEGNet universally exceeding $98\%$ area under the curve, the cross-subject evaluations expose profound inter-group disparities. The SA cohort consistently demonstrates the most superior generalization capability, achieving the highest metrics across most deep learning architectures, peaking at a $69.1\%$ area under the curve for EEGNet under the leave-one-subject-out protocol. Surprisingly, despite possessing the largest participant pool of thirty-one subjects, the HK group yields noticeably inferior cross-subject performance compared to the SA group. Meanwhile, the BF cohort, severely constrained by the smallest sample size of sixteen subjects, exhibits the most severe performance degradation, with inter-subject area under the curve dropping to as low as $48.2\%$. This nonlinear relationship between subject count and testing accuracy highlights a critical insight: raw sample size does not strictly dictate cross-subject robustness. Instead, the superior generalization of the SA group strongly suggests that the most likely primary factor is the more pronounced inherent distinction between the SA task itself and MW, which facilitates easier feature extraction. Secondarily, cohort-specific factors such as intrinsic data quality, demographic homogeneity, or distinct experimental conditions also play a vital role in forming transferable feature representations. These findings underscore the absolute necessity of identifying and addressing both task-inherent variances and dataset biases before deploying models across diverse populations.

\begin{table*}[ht]
  \centering
  \small
  \renewcommand{\arraystretch}{1.2}
  \setlength{\tabcolsep}{3.5pt}
  \caption{\small \textbf{Task 1: Group-specific Performance.} Mean AUC (\%) and BAcc (\%) $\pm$ standard deviation per site. Intra-subject results are computed from per-subject held-out evaluations within each site.}
  \label{tab:task1_groups_full}
  \begin{tabular}{@{} l l cc cc cc @{}}
    \toprule
    & & \multicolumn{2}{c}{\textbf{Intra-Subject}} & \multicolumn{2}{c}{\textbf{Inter-Subject}} & \multicolumn{2}{c}{\textbf{LOSO}} \\
    \cmidrule(lr){3-4} \cmidrule(lr){5-6} \cmidrule(lr){7-8}
    \textbf{Group} & \textbf{Model} & \textbf{AUC} & \textbf{BAcc} & \textbf{AUC} & \textbf{BAcc} & \textbf{AUC} & \textbf{BAcc} \\
    \midrule
    \multirow{5}{*}{SA ($n$=27)} 
    & ShallowConvNet & 98.4{\scriptsize $\pm$2.5} & 93.4{\scriptsize $\pm$5.5} & 61.1{\scriptsize $\pm$4.4} & 57.5{\scriptsize $\pm$4.1} & 66.9{\scriptsize $\pm$22.3} & 59.8{\scriptsize $\pm$13.8} \\
    & DeepConvNet & 96.2{\scriptsize $\pm$11.2} & 88.1{\scriptsize $\pm$14.2} & 60.6{\scriptsize $\pm$3.9} & 55.9{\scriptsize $\pm$4.1} & 68.9{\scriptsize $\pm$24.8} & 61.1{\scriptsize $\pm$18.8} \\
    & EEGNet & \textbf{99.7{\scriptsize $\pm$0.5}} & \textbf{96.6{\scriptsize $\pm$4.4}} & 60.5{\scriptsize $\pm$9.5} & 56.5{\scriptsize $\pm$6.7} & \textbf{69.1{\scriptsize $\pm$23.1}} & \textbf{61.6{\scriptsize $\pm$13.8}} \\
    & EEG-Conformer & 98.6{\scriptsize $\pm$2.6} & 93.2{\scriptsize $\pm$6.7} & \textbf{61.1{\scriptsize $\pm$4.7}} & \textbf{57.8{\scriptsize $\pm$4.1}} & 68.8{\scriptsize $\pm$26.3} & 61.2{\scriptsize $\pm$18.2} \\
    \midrule
    \multirow{5}{*}{HK ($n$=31)}
    & ShallowConvNet & 96.2{\scriptsize $\pm$5.7} & 89.4{\scriptsize $\pm$8.4} & 54.5{\scriptsize $\pm$7.2} & \textbf{54.4{\scriptsize $\pm$5.5}} & \textbf{61.2{\scriptsize $\pm$29.3}} & \textbf{57.0{\scriptsize $\pm$14.9}} \\
    & DeepConvNet & 97.4{\scriptsize $\pm$7.3} & 88.2{\scriptsize $\pm$13.0} & 54.1{\scriptsize $\pm$9.5} & 51.8{\scriptsize $\pm$8.3} & 58.3{\scriptsize $\pm$30.7} & 53.8{\scriptsize $\pm$18.6} \\
    & EEGNet & \textbf{98.8{\scriptsize $\pm$3.1}} & \textbf{94.7{\scriptsize $\pm$6.9}} & \textbf{55.5{\scriptsize $\pm$11.6}} & 52.8{\scriptsize $\pm$9.6} & 57.5{\scriptsize $\pm$31.1} & 54.2{\scriptsize $\pm$23.0} \\
    & EEG-Conformer & 96.2{\scriptsize $\pm$5.8} & 89.3{\scriptsize $\pm$8.7} & 54.0{\scriptsize $\pm$9.8} & 52.2{\scriptsize $\pm$8.4} & 60.1{\scriptsize $\pm$31.7} & 53.6{\scriptsize $\pm$19.6} \\
    \midrule
    \multirow{5}{*}{BF ($n$=16)}
    & ShallowConvNet & 97.1{\scriptsize $\pm$3.2} & 88.1{\scriptsize $\pm$7.9} & 55.2{\scriptsize $\pm$3.4} & 53.9{\scriptsize $\pm$2.5} & 51.5{\scriptsize $\pm$17.1} & 52.0{\scriptsize $\pm$7.9} \\
    & DeepConvNet & 97.7{\scriptsize $\pm$3.9} & 81.3{\scriptsize $\pm$14.9} & 50.4{\scriptsize $\pm$8.9} & 52.1{\scriptsize $\pm$4.7} & 52.4{\scriptsize $\pm$24.2} & 53.3{\scriptsize $\pm$11.5} \\
    & EEGNet & \textbf{99.1{\scriptsize $\pm$1.2}} & \textbf{94.8{\scriptsize $\pm$6.0}} & 48.2{\scriptsize $\pm$9.0} & 49.1{\scriptsize $\pm$6.3} & 49.2{\scriptsize $\pm$21.6} & 51.2{\scriptsize $\pm$8.0} \\
    & EEG-Conformer & 95.9{\scriptsize $\pm$5.4} & 87.1{\scriptsize $\pm$11.4} & \textbf{55.6{\scriptsize $\pm$5.9}} & \textbf{54.9{\scriptsize $\pm$3.1}} & \textbf{55.9{\scriptsize $\pm$20.7}} & \textbf{53.9{\scriptsize $\pm$8.4}} \\
    \bottomrule
  \end{tabular}
\end{table*}

\begin{figure}[ht]
    \centering
    \includegraphics[width=\linewidth]{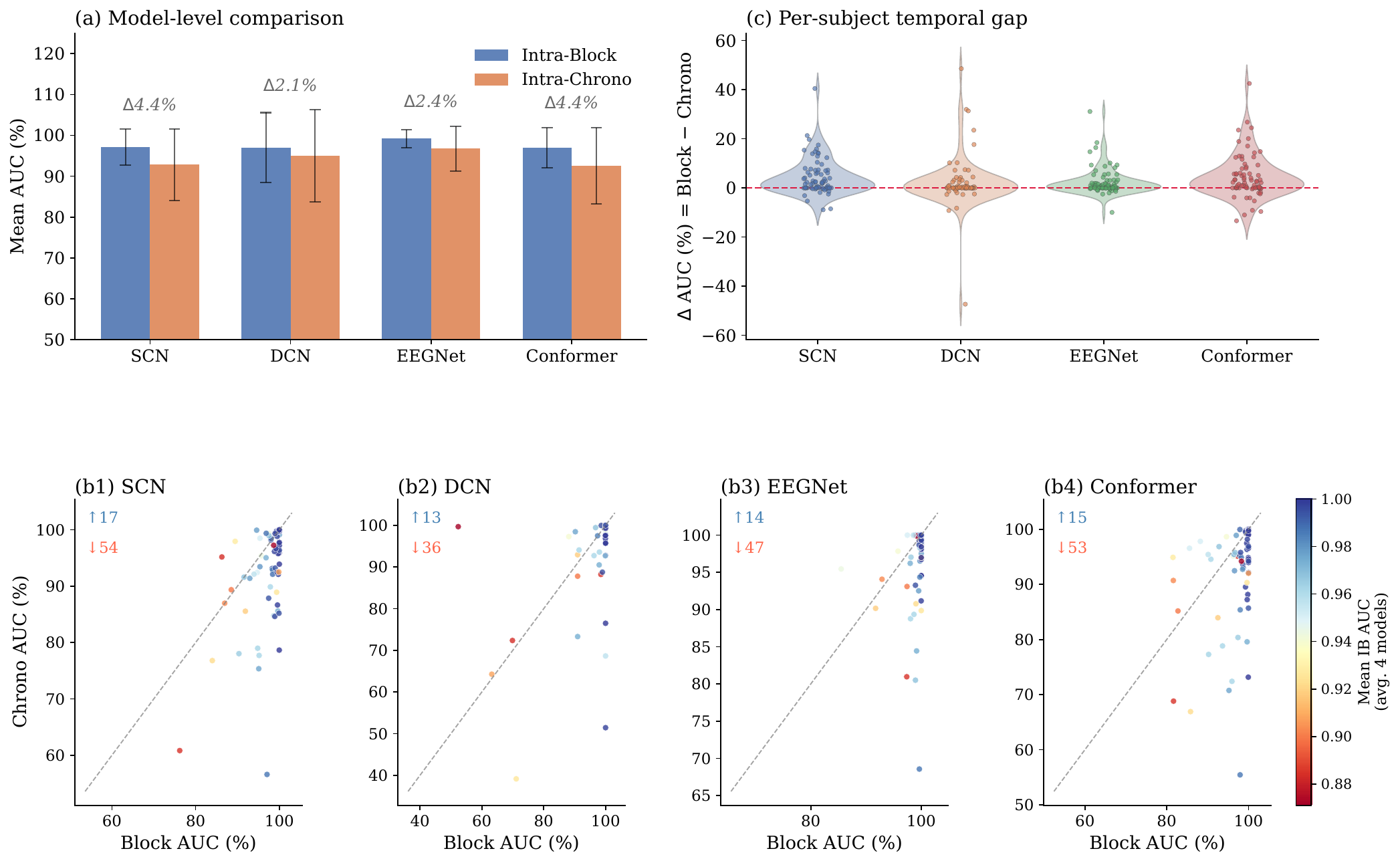}
    \caption{\textbf{Intra-Subject Evaluation Protocol Comparison: Block-Split vs.\ Chronological-Split ($N=74$)} \
\textbf{(a)}~Mean AUC~(\%) and standard deviation across all subjects for four models under two intra-subject protocols: block-wise (random fold assignment) and chrono-wise (temporally ordered folds). $\Delta$ denotes the performance gap between protocols. \
\textbf{(b1--b4)}~Per-subject scatter plots for each model; each point represents one subject, with the $x$-axis showing block-wise AUC and the $y$-axis showing chrono-wise AUC. Points below the diagonal (dashed line) indicate subjects whose performance degrades under temporal splitting. Point colour encodes each subject's mean block-wise AUC averaged across all four models (warm: lower performers; cool: higher performers). \
\textbf{(c)}~Distribution of $\Delta$AUC~$=$~Block~$-$~Chrono per subject for each model (violin: kernel density; dots: individual subjects); the red dashed line marks $\Delta = 0$. SCN:~ShallowConvNet; DCN:~DeepConvNet.}
    \label{fig:block_chrono}
\end{figure}

\paragraph{Impact Of Temporal Shifts On Evaluation Realism} The transition from block-wise to chrono-wise evaluation protocols exposes a systematic performance degradation across all evaluated architectures, with area under the curve metrics declining by $2.1\%$ to $4.4\%$ as explicitly detailed in the model-level comparison (Figure~\ref{fig:block_chrono}A). This performance gap highlights the methodological trade-offs inherent to the block-wise partitioning strategy. On the positive side, the higher accuracy achieved by block-wise splitting demonstrates that this method successfully captures the non-stationary temporal dynamics throughout the session, thereby establishing a valuable performance upper bound for within-session real-time detection applications. However, this localized randomization simultaneously introduces the risk of temporal leakage, where future signal dynamics inadvertently inform the training phase. Whether this specific leakage ultimately compromises cross-session generalization accuracy remains an open question that warrants further observation. In contrast, the chrono-wise protocol enforces strict past-to-future causality, explicitly revealing the models' vulnerability to natural signal drift over time. Furthermore, individual scatter distributions and per-subject temporal gap analyses (Figure~\ref{fig:block_chrono}B and Figure~\ref{fig:block_chrono}C) highlight that this temporal vulnerability is highly heterogeneous across subjects. While the majority of data points fall below the parity line, indicating generalized performance loss, specific individuals experience severe performance collapses whereas others maintain stable accuracy regardless of the splitting strategy. Ultimately, while the block-wise evaluation establishes a theoretical upper bound for model capacity under stationary conditions, the chrono-wise approach provides a fundamentally more rigorous and realistic assessment of cross-temporal generalization for continuous, online deployment scenarios.

\begin{figure}[ht]
    \centering
    \includegraphics[width=0.65\linewidth]{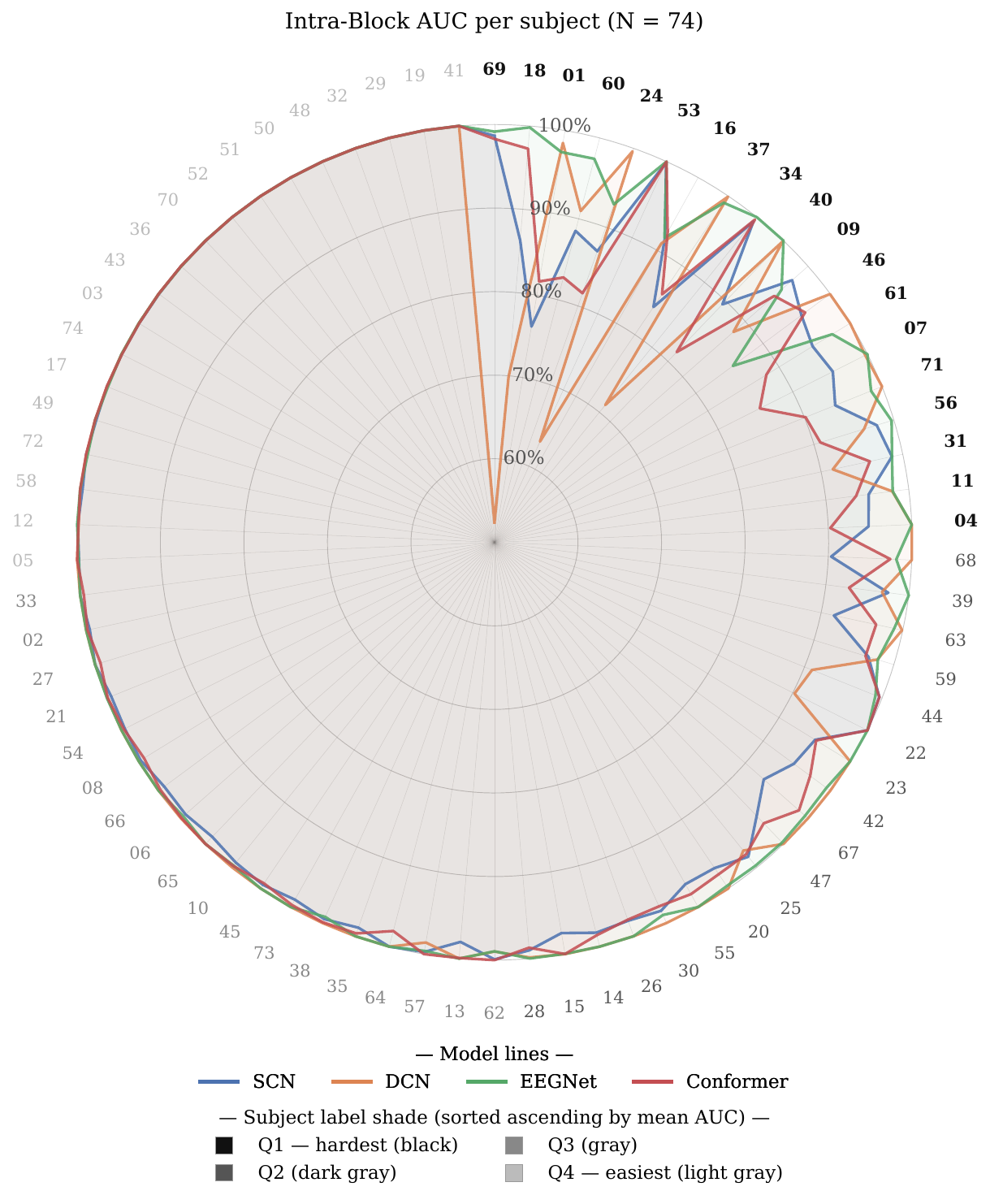}
    \caption{\small \textbf{block-wise AUC Across All Subjects and Models ($N = 74$)} Each spoke represents one subject, arranged clockwise in ascending order of mean block-wise AUC averaged across all four models; subjects in the lower-performing quartile (Q1) appear first in the sweep. The four coloured lines correspond to the four model architectures: SCN (ShallowConvNet), DCN (DeepConvNet), EEGNet, and Conformer (EEG-Conformer). The radial axis is linearly rescaled such that 50\% AUC (chance level) maps to the origin and 100\% AUC maps to the outermost ring; concentric grid circles mark 60\%, 70\%, 80\%, 90\%, and 100\%. Subject labels are shaded in greyscale according to their mean-AUC performance quartile across all models: \textbf{Q1} (black, bold) denotes subjects in the lowest quartile (${\leq}$25th percentile, most difficult to decode), \textbf{Q2} (dark grey), \textbf{Q3} (grey), and \textbf{Q4} (light grey, highest quartile, ${\geq}$75th percentile). Inward dips along a spoke indicate that at least one model achieves lower-than-average performance for that subject, highlighting subject-level heterogeneity in EEG decodability.}
    \label{fig:radar_intra}
\end{figure}

\paragraph{Analyzing Individual Variability and Cohort Biases} The primary objective of this subject-wise analysis is to evaluate intra-subject decoding performance across the dataset. While models demonstrate high state-separation accuracy for most participants, a subset of subjects remains challenging (Figure~\ref{fig:radar_intra}). This difficulty can be attributed to several factors. First, for novice practitioners engaging in silent meditation for the first time, frequent mind-wandering episodes can blur the neural distinction between meditation and resting states. Second, inherently poor signal quality or atypical EEG patterns in certain subjects complicate feature extraction. As a result, deep learning classifiers such as the Deep Convolutional Network (DCN) exhibit noticeably reduced classification accuracy on this subset.

To systematically quantify this variability, subjects are categorized into quartiles based on their average decoding accuracy across all groups: the top 25\% are defined as resilient subjects (easiest to decode), and the bottom 25\% are defined as vulnerable subjects (hardest to decode). Stratifying these distributions reveals distinct cohort differences. The SA group ($n=27$) exhibits the most robust intra-subject state separation, featuring the highest proportion of resilient subjects at 37.0\% and the lowest proportion of vulnerable subjects at 18.5\%. In contrast, the BF group ($n=16$) is heavily skewed toward the difficult end, with 68.7\% of its participants falling into the bottom half of performance (Q1 and Q2) and only 6.2\% classified as resilient. Similarly, the HK group ($n=31$) presents a higher vulnerable proportion (29.0\%) than the SA cohort. These disparities align with the initial difficulty of the respective practices for novices. The SA task uses a simple, four-syllable mantra that facilitates sustained attention during early training stages, whereas the BF task demands continuous breath monitoring and the HK task involves a longer, more complex mantra. Both BF and HK tasks are therefore more susceptible to mind-wandering during early training stages.

This subject-wise analysis provides a practical guide for future researchers using the L-FAME dataset. Rather than treating decoding failures as statistical noise obscured by aggregate metrics, this stratification highlights challenging edge cases. This benchmark offers two methodological paths for future studies. Researchers aiming to establish the upper bound of state separability can use these vulnerability metrics as an empirical exclusion criterion, selectively removing vulnerable subjects to isolate pure neural signatures of the target states. Alternatively, researchers focusing on algorithmic robustness can specifically target these vulnerable subjects. Future work should investigate the root causes of these intra-subject failures to determine whether performance degradation is driven by behavioral non-compliance (e.g., mind-wandering during silent meditation) or artifactual signal degradation. By explicitly reporting these individual differences, this analysis contextualizes the overall task difficulty and directs the community toward targeted algorithmic improvements.

% ========================================================== %

\subsection{Benchmark Task 2 Supplementary Material}
\label{app:task2}
This section investigates the neurophysiological differences among the three meditation techniques and evaluates how the six-week intervention affects their neural representations. To separate practice-induced neural adaptations from potential attrition bias, we analyze a matched paired-subject cohort. We trace the temporal evolution of representational separability using UMAP latent space visualizations and Representational Similarity Analysis (RSA), we observe that three-class classification separability improves post-training compared to pre-training from the extracted feature space, and that the distinct meditation modalities make the BF group easier to distinguish. These observations are validated with an inter-subject One-Versus-All (OvA) classification framework. The results provide statistical evidence for a clear boundary between the somatic-directed breath-focused (BF) modality and the inner-speech practices (HK and SA).

\paragraph{Controlled Longitudinal Evaluation Isolates Genuine Practice Effects From Attrition Bias} The primary evaluation of the second task initially compares the pre-intervention performance, trained on the full cohort of 74 subjects, against the post-intervention performance, trained on the 44 subjects who completed the six-week programme. Because these two conditions differ simultaneously in temporal session and subject composition, this unrestricted comparison is inherently confounded by potential attrition bias; highly engaged participants might artificially inflate post-intervention metrics independently of genuine neurophysiological changes. To eliminate this confound and isolate the true longitudinal impact, a strictly paired evaluation protocol was established. Specifically, the analysis was restricted exclusively to the subset of 44 subjects who possessed both pre- and post-intervention recordings. By evaluating their pre-intervention baseline directly against their post-intervention results using an identical inter-subject cross-validation protocol, this paired design ensures a rigorous comparison. Analyzing this strictly matched baseline reveals distinctly divergent behavioral patterns between the deep learning architectures. From Table~\ref{tab:performance_metrics_task2_prepos}, for the EEGNet model, all four evaluated metrics exhibit a monotonic increase, and crucially, the improvement persists even after the subject population is strictly held constant, with the area under the precision-recall curve rising from $42.9\%$ to $48.8\%$. The improvement within the matched cohort is consistent with increased technique-specific discriminability after training, although session-related factors and other longitudinal changes cannot be fully excluded. Conversely, the ShallowConvNet model demonstrates statistically indistinguishable performance between the controlled pre-intervention and post-intervention conditions, recording precision-recall areas of $46.9\%$ and $45.7\%$, respectively. This performance plateau suggests that the apparent improvement observed in the unrestricted comparison for ShallowConvNet is predominantly driven by demographic attrition rather than algorithmic adaptation to practice-induced neural changes. Ultimately, these complementary findings validate the critical necessity of this paired-subject dataset design, as it successfully captures authentic cognitive advancements while simultaneously preventing the misinterpretation of subject selection artifacts as algorithmic improvements.

\begin{table}[ht]
  \centering
  \small
  \caption{\small \textbf{Task 2 Performance Metrics Under Controlled Longitudinal Conditions.} The Session column denotes the temporal phase of data collection, specifically the pre-intervention (Pre) and post-intervention (Post) stages. The N value indicates the total number of subjects included in the training and evaluation cohort, where N=74 represents the full initial sample and N=44 represents the paired subset of participants who successfully completed the entire six-week meditation program.}
  \label{tab:performance_metrics_task2_prepos}
  \begin{tabular}{llcccc}
    \toprule
    \textbf{Model} & \textbf{Session(N)} & \textbf{PR-AUC} & \textbf{BAcc} & \textbf{F1} & \textbf{AUC} \\
    \midrule
    \multirow{3}{*}{EEGNet} & Pre (74) & 35.0 $\pm$ 2.9 & 35.0 $\pm$ 5.6 & 32.8 $\pm$ 5.9 & 52.4 $\pm$ 4.1 \\
     &Pre (44)& 42.9 $\pm$ 8.7 & 41.4 $\pm$ 6.6 & 40.1 $\pm$ 6.5 & 56.8 $\pm$ 5.5 \\
     & Post (44)& \textbf{48.8 $\pm$ 8.6} & \textbf{48.0 $\pm$ 7.4} & \textbf{45.0 $\pm$ 8.3 }& \textbf{64.5 $\pm$ 7.3} \\
    \midrule
    \multirow{3}{*}{ShallowConvNet} & Pre (74) & 38.0 $\pm$ 6.2 & 38.5 $\pm$ 4.4 & 33.4 $\pm$ 4.7 & 55.4 $\pm$ 6.4 \\
     & Pre (44) & \textbf{46.9 $\pm$ 4.5} & 43.9 $\pm$ 5.1 & 40.5 $\pm$ 4.5 & 59.2 $\pm$ 6.3 \\
     & Post (44) & 45.7 $\pm$ 9.7 & \textbf{44.7 $\pm$ 7.1} & \textbf{42.9 $\pm$ 8.0} & \textbf{60.0 $\pm$ 6.7} \\
    \bottomrule
  \end{tabular}
\end{table}

\paragraph{Longitudinal Intervention Enhances Representational Separability} 

\begin{figure}[ht]
    \centering
    \includegraphics[width=0.9\linewidth]{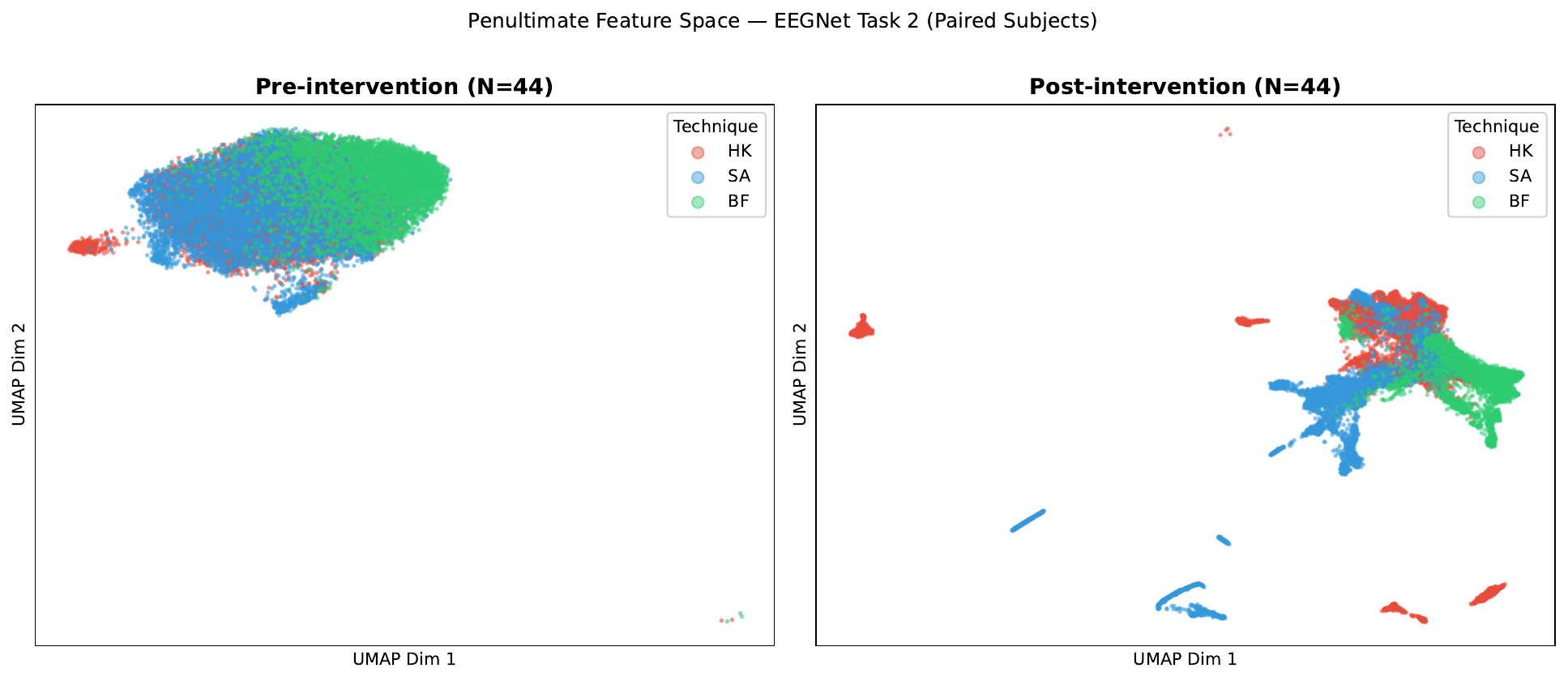}
    \caption{\small \textbf{UMAP visualization of EEGNet penultimate-layer representations for the 44 paired subjects before (left) and after (right) the six-week meditation intervention, with separate models trained for before and after intervention.} Each point represents one 4-second EEG window, colored by meditation technique (HK: red, SA: blue, BF: green). UMAP is fitted jointly on the combined features so that both panels share a common coordinate system. 
    Note that some of the data points (e.g., for HK) are covered by other classes' points.}
    \label{fig:umap}
\end{figure}

To rigorously determine whether the observed improvements in technique classification accuracy stem from sample size variations or a genuine longitudinal strengthening of technique-specific neural patterns, a paired analysis was conducted on forty-four subjects possessing both pre-intervention and post-intervention recordings. Specifically, features are extracted from the final hidden layer of an EEGNet model trained on post-intervention data and applied identically to both sessions. This strict control ensures that the resulting embedding space reflects actual changes in the neural data rather than shifting model weights. By fitting UMAP jointly on the combined pre- and post-session features ($n_{\mathrm{neighbors}}=30$, $\mathrm{min\_dist}=0.15$) to establish a common coordinate system, the visualization reveals a striking representational shift. In the pre-intervention phase, the feature embeddings for the three meditation techniques remain largely entangled, indicating relatively weak class-specific neural signatures prior to structured practice. Specifically, while the HK (red cluster) and SA (blue cluster) cohorts exhibit substantial overlap, the BF (green cluster) group already begins to demonstrate an incipient separation, although with some residual intersection. In contrast, the post-intervention latent space displays a markedly more dispersed distribution with clearer categorical boundaries (Figure~\ref{fig:umap}). Although a minor overlap remains visible between the HK and SA groups, the BF group becomes clearly isolated into a highly distinct cluster. This progressive separation trajectory highlights the distinct underlying cognitive mechanisms of these practices: the BF technique, predominantly involving somatic attention directed toward respiration, elicits unique neurophysiological features that become readily separable from the inner-speech-focused HK and SA techniques even after short-term training.

\begin{figure}[ht]
    \centering
    \includegraphics[width=0.9\linewidth]{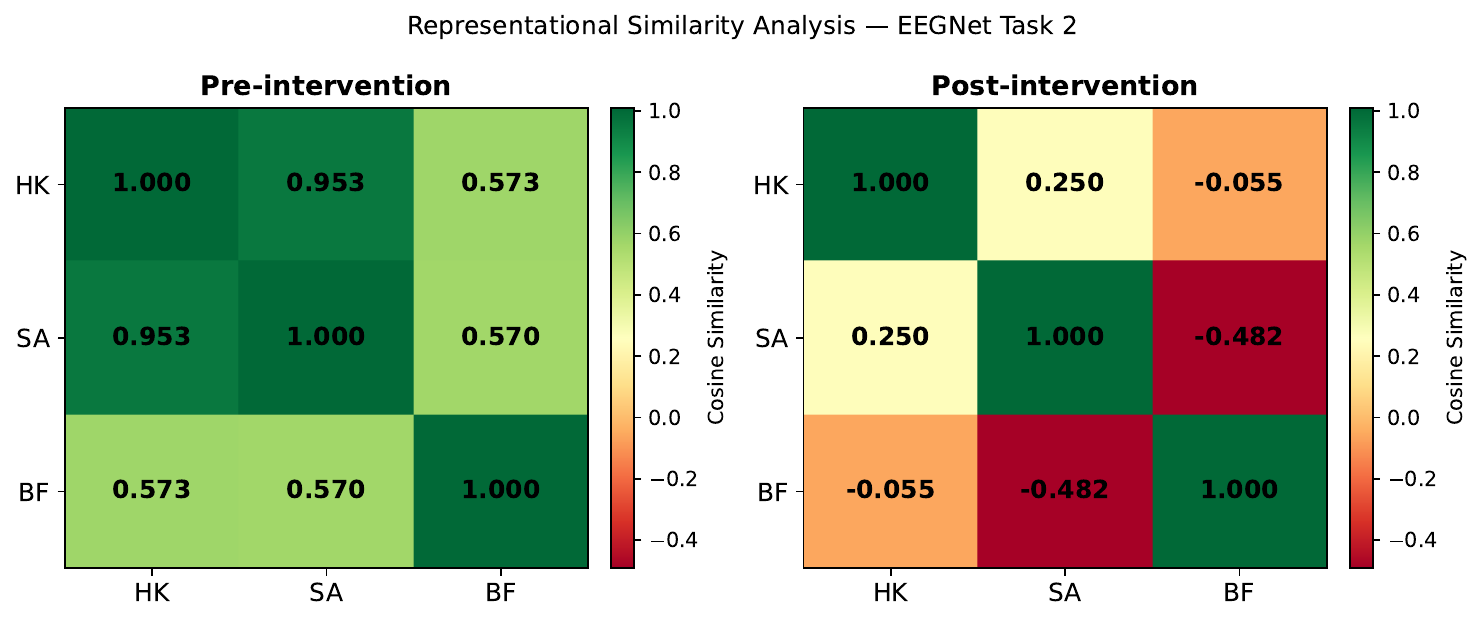}
    \caption{\small \textbf{Pre-intervention and post-intervention representational similarity analysis (RSA) matrices for EEGNet} illustrate the cosine similarity between the neural representation centroids of three techniques (HK, SA, and BF), where both numerical values and color gradients denote the corresponding similarity scores.}
    \label{fig:rsa}
\end{figure}

To formally validate the visual observations, representational similarity analysis is performed by averaging the high-dimensional feature embeddings within each category to establish class centroids, followed by computing the pairwise cosine similarities among these three centers (Figure~\ref{fig:rsa}). The resulting similarity matrices strictly corroborate the progressive separation trajectory. In the pre-intervention phase, the neurophysiological representations exhibit severe entanglement. The inner-speech-focused HK and SA cohorts share a remarkably high cosine similarity of 0.953 , while the BF group also maintains substantial positive correlations with HK at 0.573 and SA at 0.570. After the intervention, the representational landscape undergoes a profound transformation. The similarity between the HK and SA centroids decreases to 0.250, reflecting a measurable divergence despite their shared cognitive foundation. More importantly, the BF centroid becomes structurally independent, exhibiting near-orthogonal or inverse relationships with HK at $-0.055$ and SA at $-0.482$. This stark quantitative shift from high positive correlation to orthogonal or negative similarity rigorously confirms that the somatic-directed BF technique cultivates a profoundly distinct neural signature compared to the inner-speech modalities following structured practice.

\begin{figure}[ht]
    \centering
    \includegraphics[width=0.95\linewidth]{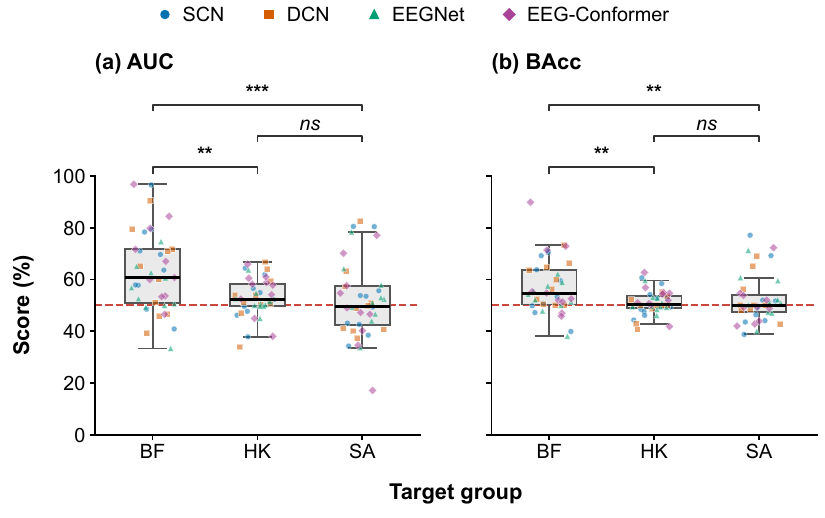}
    \caption{\small
    \textbf{Task~2 one-vs-all classification performance by target group.}
    \textbf{(a)}~AUC~(\%) and \textbf{(b)}~balanced accuracy (BAcc, \%)
    for each target group (BF, HK, SA) under the inter-subject
    one-vs-all (OvA) protocol.
    Boxes span the interquartile range (IQR), thick black lines indicate
    medians, and whiskers extend to the most extreme observations within
    $1.5\times\mathrm{IQR}$ of the box edges.
    All individual runs, including outliers, are overlaid with horizontal jitter.
    Colours and marker shapes identify SCN (ShallowConvNet; circles),
    DCN (DeepConvNet; squares), EEGNet (triangles), and EEG-Conformer (diamonds).
    Each group contains 40 run-level observations
    ($4$ architectures $\times$ $5$ folds $\times$ $2$ sessions).
    The red dashed line marks the 50\% chance reference for both metrics.
    Brackets indicate exploratory one-sided Mann--Whitney~$U$ comparisons
    (BF~$>$~HK and BF~$>$~SA), using unadjusted $p$-values:
    $^{***}p < 0.001$, $^{**}p < 0.01$, $^{*}p < 0.05$, and
    \textit{ns} for $p \geq 0.05$.
    These pooled run-level tests do not account for dependence across
    folds, sessions, or architectures.}
    \label{fig:task2_ova_group}
\end{figure}

\paragraph{Quantitative Validation Of Technique Separability Via One-Versus-All Classification} To quantitatively validate the geometric separability observed in the prior representations, the three-way technique classification was decomposed into three independent binary classifiers via an One-Versus-All (OvA) approach. For each OVA model, the labeling scheme follows a strict binary dichotomy: data from the designated target group (e.g., HK, SA, or BF) are assigned to the positive class, while data from the remaining two groups are aggregated into the negative class. A rigorous 5-fold inter-subject cross-validation strategy is strictly enforced to evaluate genuine generalization capabilities and prevent data leakage. Specifically, the entire subject pool is partitioned into five mutually exclusive subsets. In each iteration of the cross-validation process, four subsets (comprising 80\% of the individuals) are allocated to the training set to optimize the classifiers, while the single remaining subset (comprising 20\% of the individuals) is exclusively held out for the testing phase. By iteratively ensuring that the training and testing sets contain completely non-overlapping subject pools across all five folds, this inter-subject training paradigm guarantees that the models learn universal, technique-specific cognitive traits rather than overfitting to individual-level physiological artifacts. Because of this structural skew, the binary AUC and BAcc serve as the most rigorous and sensitive metrics for evaluation. The resulting performance distributions across all evaluated architectures reveal a stark asymmetry in class distinctness. When the BF cohort is isolated as the positive target class, models achieve the highest decoding metrics, distinctly surpassing theoretical chance baselines and demonstrating a statistically significant performance advantage over the alternative techniques. Conversely, when either the HK or SA cohort is designated as the singular target class, the classification performance collapses to near-chance levels. This quantitative discrepancy strictly corroborates the previously discussed representational structures. It confirms that the BF meditation technique, which relies heavily on somatic attention toward respiration, produces a highly distinct and separable neurophysiological pattern. In contrast, the HK and SA techniques, which share underlying cognitive mechanisms related to inner speech, exhibit high mutual confusability and poor independent separability. Consequently, the elevated classification efficacy when isolating the BF group demonstrates that the observed representational shifts are driven by genuine, task-specific cognitive divergences, underscoring the profound neurophysiological distinction between somatic-focused and inner-speech-focused meditation practices. As depicted in the performance boxplots (Figure~\ref{fig:task2_ova_group}), statistical test confirm a statistically significant advantage when the BF cohort is designated as the target class, compared to the two inner-speech-focused groups. Notably, the individual data points overlaid on these boxplots represent distinct performance observations comprehensively sampled across four model architectures, both pre- and post-intervention sessions, and five cross-validation folds. These exploratory comparisons suggest greater one-vs-all discriminability for BF than for either mantra-based technique, although the pooled run-level tests do not account for dependencies across folds, sessions, or architectures.

% ========================================================== %
\subsection{Benchmark Task 3 Supplementary Material}
\label{app:task3}

Full fine-tuning updated all learnable model parameters. In contrast, BN-and-head fine-tuning updated only the BatchNorm affine parameters and the final linear classification head. For Conformer, all Transformer parameters, including LayerNorm affine parameters, remained frozen under BN-and-head fine-tuning.

Both strategies used 30 adaptation epochs and a batch size of four. The learning rate was set to \(5\times10^{-5}\) for EEGNet, ShallowConvNet, and DeepConvNet, and \(10^{-5}\) for Conformer.

During adaptation on the calibration set, BN-and-head adaptation used Adam with zero weight decay and a cross-entropy loss. Training samples were drawn with replacement using inverse-class-frequency weights. During adaptation, the models remained in training mode, allowing BatchNorm running statistics to update and Dropout to remain active. The model obtained after the final adaptation epoch was evaluated on the held-out post-intervention test set, without validation-based early stopping or checkpoint selection.

\begin{figure}[ht]
    \centering
    \begin{minipage}{0.55\textwidth}
        \centering
        \includegraphics[width=\linewidth]{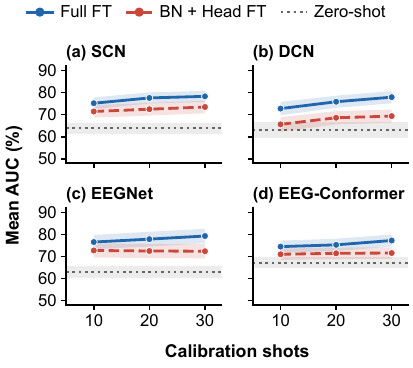}
        \caption{\small
        \textbf{Few-Shot Adaptation Learning Curves for Task~3
        ($N = 44$ subjects).}
        Each panel shows mean AUC~(\%) at 10, 20, and 30 calibration
        shots for one architecture:
        \textbf{(a)}~SCN (ShallowConvNet),
        \textbf{(b)}~DCN (DeepConvNet),
        \textbf{(c)}~EEGNet, and
        \textbf{(d)}~EEG-Conformer.
        Solid blue lines denote full fine-tuning (Full FT), which
        updates all model parameters. Dashed red lines denote
        BN \& linear head fine-tuning (BN + Head FT), which updates
        the BatchNorm layers and the linear classification head
        while keeping the remaining model parameters frozen.
        Shaded bands represent $\pm 1$ standard error of the mean
        across subjects. The grey dotted line and shaded region
        indicate the zero-shot mean and $\pm 1$ standard error.
        Full fine-tuning achieves higher mean AUC than
        BN \& linear head fine-tuning for all four architectures
        at each evaluated shot count.}
        \label{fig:task3_learning_curve}
    \end{minipage}\hfill
    \begin{minipage}{0.4\textwidth}
        \centering
        \includegraphics[width=\linewidth]{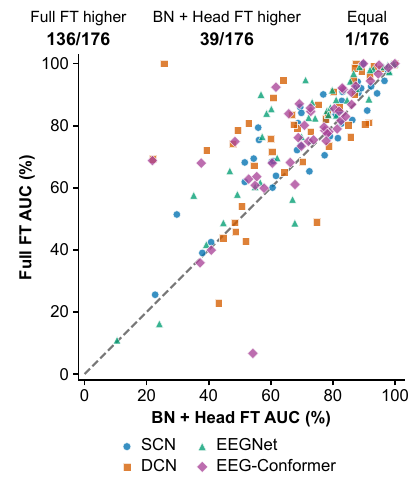}
        \caption{\small
        \textbf{Per-Subject Comparison of Full Fine-Tuning and
        BN \& Linear Head Fine-Tuning at 30 Shots
        ($N = 44$ subjects).}
        Each point represents one subject--model pair
        ($44 \times 4 = 176$ pairs).
        The $x$-axis shows AUC~(\%) under BN \& linear head
        fine-tuning (BN + Head FT), and the $y$-axis shows
        AUC~(\%) under full fine-tuning (Full FT).
        Colours identify the four architectures:
        SCN (ShallowConvNet), DCN (DeepConvNet), EEGNet,
        and EEG-Conformer.
        The dashed diagonal indicates equal performance ($y=x$).
        Points above the diagonal favour full fine-tuning,
        whereas points below favour BN \& linear head fine-tuning.
        The annotations report the number of pairs favouring
        each strategy. Most pairs favour full fine-tuning,
        although the relative benefit varies across subjects
        and architectures.}
        \label{fig:task3_paired_scatter}
    \end{minipage}
\end{figure}

\paragraph{Monotonic Performance Gains And The Superiority Of Full Fine-Tuning} The evaluation of cross-session generalization, where models trained on pre-intervention data are tested on post-intervention sessions, reveals critical insights into few-shot adaptation strategies (Figure~\ref{fig:task3_learning_curve}). As demonstrated by the learning trajectories across all evaluated architectures, including ShallowConvNet, DeepConvNet, EEGNet, and EEG-Conformer, the area under the curve exhibits a consistent monotonic increase as the number of available training shots grows from 10 to 30 shots. Notably, the zero-shot reference baseline remains substantially lower than both adaptation curves, indicating that pre-intervention representations alone are insufficient for optimal cross-session transfer due to natural temporal distribution shifts. However, introducing even a minimal 10-shot calibration yields significant performance recoveries. In comparing adaptation paradigms, full fine-tuning consistently outperforms BN \& linear head fine-tuning across all shot capacities and model architectures. This persistent gap highlights that updating the entire hierarchical weight structure is fundamentally necessary to capture the subtle neurophysiological changes induced over time, whereas restricting updates to the final classification layer via BN \& linear head fine-tuning limits the capacity of the model to adapt to newly emerged, technique-specific data distributions.

\paragraph{Universal Subject-Level Efficacy And Architecture-Agnostic Adaptation} The performance advantage of full fine-tuning over BN \& linear head fine-tuning extends beyond aggregate metrics, demonstrating strong consistency at the individual subject level (Figure~\ref{fig:task3_paired_scatter}). An analysis of the paired scatter distribution under the 30-shot condition reveals that the vast majority of the 176 measurements (derived from 44 subjects evaluated across 4 models) lie above the parity line. The overall effectiveness of full fine-tuning suggests that updating the feature extractor is generally necessary to capture the shifts in the temporal features of EEG signals induced by short-term training interventions. However, the minority of instances where BN \& linear head fine-tuning outperforms full fine-tuning warrants further explanation. Excluding extreme cases where both evaluation methods exhibit poor performance, which is likely attributable to degraded signal quality, the majority of data points favoring BN \& linear head fine-tuning cluster closely around the parity line. This observation indicates that short-term meditation training did not induce substantial changes in the EEG signals for these specific subjects. Under such conditions, the initial feature representations remain adequate, allowing BN \& linear head fine-tuning to achieve competitive performance of full fine-tuning while avoiding the potential overfitting risks associated with full network updates.

%%%%%%%%%%%%%%%%%%%%%%%%%%%%%%%%%%%%%%%%%%%%%%%%%%%%%%%%%%%%

%\input{checklist.tex}

\end{document}